\documentclass[]{aastex}

\usepackage{emulateapj5,ifthen}
\shorttitle{$K$-band Properties of Galaxy Clusters}
\shortauthors{Lin, Mohr \& Stanford}

\def\Om{$\Omega_M$\ }

\def\xray{\hbox{X--ray} }

\def\kev{{\rm\ keV}}
\def\mlr{\hbox{mass--to--light ratio}\ }

\def\mlre{\hbox{mass--to--light ratio}}

\newcommand{\figtype}{EPS}
\def\myputfigure#1#2#3#4#5%
{\vskip#5pt\makebox[0pt]{\hskip#2in
\includegraphics[width=#3\textwidth]{#1}}\vskip#4pt\hfill}
\newenvironment{inlinefigure}{
\def\@captype{figure}
\noindent\begin{minipage}{0.999\linewidth}\begin{center}}
{\end{center}\end{minipage}\smallskip}

\begin{document}

\slugcomment{Accepted for publication in ApJ}

\title{$K$-band Properties of Galaxy Clusters and Groups: \\ Luminosity Function, Radial Distribution and Halo Occupation Number}

\author{Yen-Ting Lin\altaffilmark{1}, Joseph J. Mohr\altaffilmark{1,2}
and S. Adam Stanford\altaffilmark{3,4}}
\altaffiltext{1}{Department of Astronomy, University of Illinois,
Urbana, IL 61801; ylin2@astro.uiuc.edu}
\altaffiltext{2}{Department of Physics, University of Illinois,
Urbana, IL 61801; jmohr@uiuc.edu}
\altaffiltext{3}{Physics Department, University of California,
Davis, CA 95616; adam@igpp.ucllnl.org}
\altaffiltext{4}{Institute of Geophysics and Planetary Physics, Lawrence 
Livermore National Laboratory, Livermore, CA 94551}

\begin{abstract}

We explore the near-infrared (NIR) $K$-band properties of galaxies within 93
galaxy clusters and groups using data from the Two Micron All Sky Survey 
(2MASS).  We use X-ray properties of these clusters to pinpoint cluster centers 
and estimate cluster masses.
By stacking all these systems, we study the shape of the cluster luminosity
function and the galaxy distribution within the clusters. We find that the 
galaxy profile is well described by the NFW profile
with a concentration parameter $c \sim 3$, with no evidence for cluster mass
dependence of the concentration. Using this sample, whose masses span 
the range from $3\times10^{13}M_\odot$ to $2\times10^{15}M_\odot$, we confirm 
the existence of a tight correlation between total galaxy NIR luminosity and 
cluster binding mass, which indicates that NIR light can serve as a cluster mass
indicator. 
From the observed galaxy profile, together with cluster 
mass profile measurements from the literature,
we find that the \mlr is a weakly decreasing function of cluster radius, and 
that it increases with cluster mass. We also derive the mean number of galaxies 
within halos of a given mass, the halo occupation number. We 
find that the mean number scales as $N\propto M^{0.84\pm0.04}$ for galaxies
brighter than $M_K=-21$, 
indicating high mass clusters have fewer galaxies per unit mass than low 
mass clusters.  Using published observations at high redshift, we show 
that higher redshift clusters have higher mean occupation number than 
nearby systems of the same mass.  By comparing the luminosity function 
and radial distribution of galaxies in low mass and high mass clusters, 
we show that there is a marked decrease in the number density of galaxies 
fainter than $M_*$ as one moves to higher mass clusters; in addition, 
extremely luminous galaxies are more probable in high mass clusters.  
We explore several processes-- including tidal interactions and merging-- 
as a way of explaining the variation in galaxy population with cluster mass.
\end{abstract}


\section{Introduction}
\label{sec:intro}

Understanding galaxy formation is one of the most outstanding
challenges in cosmology. The development of both semianalytic
\citep[e.g.][among others]{swhite78,kauffmann93,cole00} and numerical
\citep[e.g.][]{kauffmann99,springel01b} modeling have
enjoyed tremendous success, in the sense that those models are able to
match several observed properties, such as the galaxy luminosity
function (LF), the morphological mix, the color, the mass-to-light ratio of the
galaxies \citep{cole00}. But as the quality of the observational constraints
improves, we can expect additional theoretical challenges 
\citep[e.g.][]{benson03}.

The clustering properties of dark matter and galaxies pose another
tough task, for both theorists and observers.  The recent advent of
the so-called halo model has introduced an important new tool on this
subject \citep[e.g.][]{seljak00,peacock00,scoccimarro01}.  The
essential ingredients of this model include a description of halo
abundance as a function of cosmic time and halo mass (the mass
function, e.g. \citealt{press74,sheth99, jenkins01}), a model for halo
structure \citep[e.g.][]{navarro97,moore98}, and a prescription for
the bias of haloes \citep[e.g.][]{mo96,sheth99}.

The halo occupation distribution (HOD) is a powerful tool which links
the physics of galaxy formation with the clustering of dark matter and
galaxies \citep{benson00,kauffmann99,berlind02,berlind03,kravtsov03}.
It assumes that the evolution and clustering of haloes are determined
by the underlying cosmology, and that the physics that governs 
galaxy formation specifies the way galaxies populate the
haloes.  The calculations of various power spectra of dark matter and
galaxies are a natural outcome.

The important ingredients within the HOD framework are
the mean number of galaxies per halo $N$ as a function of halo mass,
the probability distribution that a halo of mass
$M$ contains $N$ galaxies $P(N|M)$, and the relative distribution
(both spatial and velocity) of galaxies and dark matter within
haloes \citep{berlind02}.

Here we aim to provide observational constraints on the HOD,
based on our study of 93 clusters and groups using data from the Two
Micron All-Sky Survey (2MASS, \citealt{jarrett00}). Using X-ray properties of 
these clusters to define the cluster center and estimate the cluster binding 
mass, we determine the mean halo occupation number $N$ as a function of mass 
from $\sim 3 \times 10^{13} M_\odot$ to $\sim 2 \times 10^{15} M_\odot$ and also
investigate the galaxy distribution and luminosity function within the clusters.
We discuss the bearing of the $N$--$M$ relation on the hierarchical structure
formation paradigm.

Part of our analysis is built on the tools that we develop in an
earlier paper \citep[][hereafter paper I]{lin03b}, where we examined
the near-infrared (NIR) galaxy luminosity--cluster binding mass
correlation ($L$--$M$ relation) for a sample of 27 nearby clusters, using the second
release of the 2MASS data.  Here, for a much larger sample, we will 
study the galaxy distribution within the clusters and the faint-end
shape of the cluster luminosity function, solve the $K_s$-band luminosity 
functions for individual clusters, and derive the total light and galaxy number 
within the virial radius as a function of cluster binding mass.  Our findings 
provide some constraints on cluster evolution scenarios.

In \S\ref{sec:analysis} we briefly describe the technique developed
in paper I for examination of the cluster NIR $L$--$M$ relation. Next in
\S\ref{sec:stack} we begin our analysis with two fundamental
properties of galaxy clusters: the luminosity function (LF) and the
spatial distribution of the member galaxies.  With constraints on
galaxy distributions both in real space and in luminosity space
derived directly from the data, we proceed to calculate the $L$--$M$
relation, point out the importance of the contributions from the
brightest cluster galaxies (BCGs) (\S\ref{sec:lm}), and examine the
mean halo occupation number (\S\ref{sec:hod}). 
We investigate the possible mechanisms that are responsible for the
observed behavior of the halo occupation distribution in
\S\ref{sec:further}.  Possible systematics that may affect our results are
discussed in \S\ref{sec:system}. Finally, in \S\ref{sec:summary}, 
we summarize our results. The Appendix provides some further tests of the
robustness of our analysis.

Throughout the paper we assume the density parameters for the matter and the
cosmological constant to be $\Omega_M = 0.3$, $\Omega_\Lambda = 0.7$, 
respectively, and the Hubble parameter to be
$H_0=70\,h_{70}$~km~s$^{-1}$~Mpc$^{-1}$.

\section{Analysis Overview}
\label{sec:analysis}

Our analysis begins with identifying a sample of clusters which have
reliable mass and cluster center estimates. Based on this
information we extract galaxies from the 2MASS extended source catalog
that lie within the virial radius for each cluster.  The
number and light of member galaxies in each cluster are estimated in a
statistical sense.  We first stack the radial distributions of galaxies
to study the galaxy distribution.
We then build a composite luminosity function from
all clusters in the sample, which allows us to examine the faint end
slope of the luminosity function (LF).  With this information, we
can derive the luminosity functions of individual clusters in a 
self-contained way.  Equipped with these tools, we are
able to address issues relating to the halo occupation distribution
(\S\ref{sec:lmhod}), and the possible transformation of galaxy
populations as low mass clusters accrete, merge and grow into high mass 
clusters (\S\ref{sec:further}).

X--ray observations provide robust estimates of cluster binding
mass and the position of the center. In particular, the tight
correlation between emission-weighted mean \xray temperature $T_X$ of
the intracluster medium (ICM) and cluster total mass indicates that
$T_X$ serves as a good proxy for mass
\citep[e.g.][]{evrard96,finoguenov01}.  We therefore assemble our
sample from several existing \xray cluster samples
\citep[][]{mohr99,reiprich02,finoguenov01,sanderson03,david93,ohara03}, and
use {\it measured} $T_X$ (i.e. not inferred from \xray luminosity or
other observables) to estimate cluster mass. The redshift information
is obtained from NED and/or SIMBAD, and the above catalogs.  The
emission-weighted mean \xray temperature is also taken from the
literature cited above.
The \xray centers are either estimated from archival ROSAT images, or taken from
literature \citep{reiprich02,boehringer00,ebeling96,ebeling98}.

We only consider systems which do not have nearby neighbors: we
eliminate the systems that have clusters or groups with measured
redshifts (i.e. confirmed identity) lying within a circle of 
$2 \sim 3$ virial radii projected on
the sky. We further remove the systems for which the signal-to-noise
is too low, i.e. the ratio of the number of galaxies within the virial
radius to the estimated statistical background/foreground galaxy
number (see below).  Finally, to reduce any confusion from stellar
objects in the 2MASS catalog, we select systems at galactic latitude
$|b|>10^\circ$.  With these criteria we end up with a sample of 93
clusters and groups (hereafter we often refer to all systems in our
sample as clusters, for simplicity). The \xray temperature in our
sample ranges from 0.8 to $\sim 12$ keV, with a redshift range $0.01 \lesssim
z \lesssim 0.09$.

Given $T_X$ of the clusters, we use the mass--temperature ($M$-$T_X$) 
relation provided by \citet{finoguenov01}
\begin{equation}
\label{eq:mt}
M_{500} = 2.55^{+0.29}_{-0.25}\,10^{13} {M_\odot\over h_{70}}\,
    \left( {T_X \over 1\kev} \right)^{1.58^{+0.06}_{-0.07}},
\end{equation}
to obtain $M_{500}\equiv (4\pi/3)500\rho_c r_{500}^3$, the mass
enclosed by $r_{500}$, within which the mean overdensity is 500
times of the critical density of the universe $\rho_c$. The virial radius
$r_{200}$ is converted from $r_{500}$ using the profile proposed by
\citet[][hereafter NFW]{navarro97} with concentration $c_{dm} = 5$
\citep[e.g.][]{vandermarel00,biviano03}.  We then search the 2MASS
all-sky release archive to collect galaxies that lie within $r_{200}$
and are brighter than the completeness limit $K_{s,lim} = 13.5$ 
(hereafter we denote $K_s$ as $K$ for simplicity). We use
the ``20 mag/square arcsec isophotal fiducial elliptical aperture
magnitude'' of the 2MASS final release.  We estimate the total
magnitudes by subtracting 0.2 mag from the isophotal magnitudes
\citep{kochanek03}.  The number and brightnesses of foreground/background
galaxies that lie within the cluster virial region is estimated from
the $\log N - \log S$ relation derived from the 2MASS all-sky data
(T. Jarrett 2003, private communication). 
In addition to the Poisson uncertainty, the contribution from background galaxy
clustering is also included in our error analysis in \S~\ref{sec:lmhod} (e.g. 
\citealt{peebles80}, see the Appendix); although this term is ignored in
\S\S~\ref{sec:stack} \& \ref{sec:further}, it does not affect our analysis or results (see the discussion in
\S\ref{sec:clf}).
The background subtracted
galaxy number and luminosity, $N_{obs}$ \& $L_{obs}$, enable us to solve for
the cluster LF,
which is assumed to be of the \citet{schechter76} form:
\begin{equation}
\label{eq:schechter}
\phi(L)\,dL = \phi_* \left( {L \over L_*} \right)^\alpha\, {\rm e}^{-L/L_*} d\left({L \over L_*}\right),
\end{equation}
where the parameters $L_*$, $\phi_*$ and $\alpha$ are the
characteristic luminosity and number densities, and the faint-end
power law index, respectively.  We seek constraints on $\alpha$
directly from our data by stacking the luminosity functions of
galaxies in all 93 clusters (\S\ref{sec:clf}). The remaining two
parameters are then determined from the two observables. We note that
the BCGs usually do not conform to the cluster LF, and we treat them
separately when determining the number count and luminosity required
to solve for the LF parameters.  We identify BCGs as the brightest galaxies
within the cluster virial radius, and we obtain their redshifts from NED to 
ensure cluster membership. A further investigation on the BCG properties will
be presented in a separate paper (Lin and Mohr 2004, in preparation, hereafter
paper III).
Once $L_*$ \& $\phi_*$ are found, we
integrate the LF to obtain the total luminosity and number of galaxies
(now including the BCGs) to a limiting magnitude $M_{K,low}=-21$.
Our approach is described in greater detail in \S 2 of paper I.

\section{Basic Properties of Stacked Clusters}
\label{sec:stack}

In this section we first study the galaxy distribution in clusters by their
surface density profile (\S\ref{sec:profile}), then in
\S\ref{sec:clf} examine the composite cluster LF (especially the 
faint-end behavior). These are needed for the next section, where we solve for 
the LF for each cluster and find the expected galaxy (subhalo) number.

\subsection{Galaxy Surface Density Profile}
\label{sec:profile}

An important aspect of the HOD formalism is the galaxy distribution
within dark matter haloes. Different distributions will result in
different clustering statistics (e.g. two-point correlation function)
at small scales \citep{seljak00,peacock00,berlind02}. Here we study
the galaxy distribution within our cluster sample by stacking the
projected galaxy distributions.

Large $N$-body simulations suggest that dark matter has a ``universal'' density
profile (NFW, \citealt{navarro03}).  The profile is
characterized by a scale radius $r_s \equiv r_{200}/c_{dm}$, where
$c_{dm}$ is the concentration parameter. The overall normalization of
the density profile is characterized by the halo formation epoch; when
the normalization is scaled out, haloes of different mass are expected
to have the same structure. We therefore stack the clusters with
radial distance scaled by their virial radii $r_{200}$.

Because of the different nature of galaxies and dark matter,
it is not necessary that both follow the same distribution.
We express the 3D number density profile as $n(x) = n_0
x^{-1} (1+x)^{-2}$, where $n_0$ is the normalization, and $x = c_g
r/r_{200}$. Note that we allow the galaxies to have concentrations
different from that of dark matter. The surface density is then an 
integral of the 3D profile
$\Sigma(x)=2n_0 r_s \int_0^{\pi/2} \cos \theta (\cos \theta+x)^{-2}
d\theta$ (see e.g. \citealt{bartelmann96}).

We assume the observed surface density is composed of the projection of the 
cluster profile and a constant background.  The best-fit surface density profile
is determined by a maximum likelihood method. Specifically, we minimize the 
quantity
\begin{equation}
-\log \mathcal{L} = -\sum_i^{N_{obs}} \log \left( \frac{\Sigma(x_i) x_i}{N_{tot}} \right) 
	- \log G(N_{obs},N_{tot}),
\end{equation}
where $N_{obs}$ is the total number of galaxies in the stacked cluster, $x_i$
is the radial distance of the $i$-th galaxy, $N_{tot}$ is the model prediction 
of the total number of galaxies (member galaxy number $N_{cls}$ plus background)
as a function of $n_0$ \& $c_g$, and $G$ is a Gaussian with mean of $N_{tot}$ 
and standard deviation $\sqrt{N_{tot}}$. In the fitting routine we allow the 
parameters $N_{cls}$ and $c_g$ to vary, while $N_{tot}$ is constrained by the 
Gaussian kernel to reproduce the observed total galaxy number $N_{obs}$.

Fig~\ref{fig:surfden} shows the radial distribution of galaxies in the stacked
cluster from $\sim 0.02\, r_{200}$ out to $2.5\, r_{200}$, as well as the 
best-fit NFW profile (which is 
obtained using the data out to $r_{200}$ only). Out to the virial radius,
the stacked cluster contains 6608 galaxies (the BCGs are excluded), of which 
1467 are estimated to be background.  It is interesting to see how 
well the simple NFW model fit works out to $r \gtrsim r_{200}$, given that we
only use a statistical background subtraction method, and the fact that we do
{\it not} pre-select the clusters to be circular/regular in shape or in 
dynamical 
equilibrium.  The {\it galaxy} concentration is $c_g=2.90_{-0.22}^{+0.21}$, 
where the uncertainties are determined from the relation $\Delta \chi^2 = 
-2\Delta \log \mathcal{L}$ with two degrees of freedom. We also notice the unit 
of the surface density is ``number per virial area''.

\begin{inlinefigure}
   \ifthenelse{\equal{\figtype}{EPS}}{
   \begin{center}
   \epsfxsize=8.cm
   \begin{minipage}{\epsfxsize}\epsffile{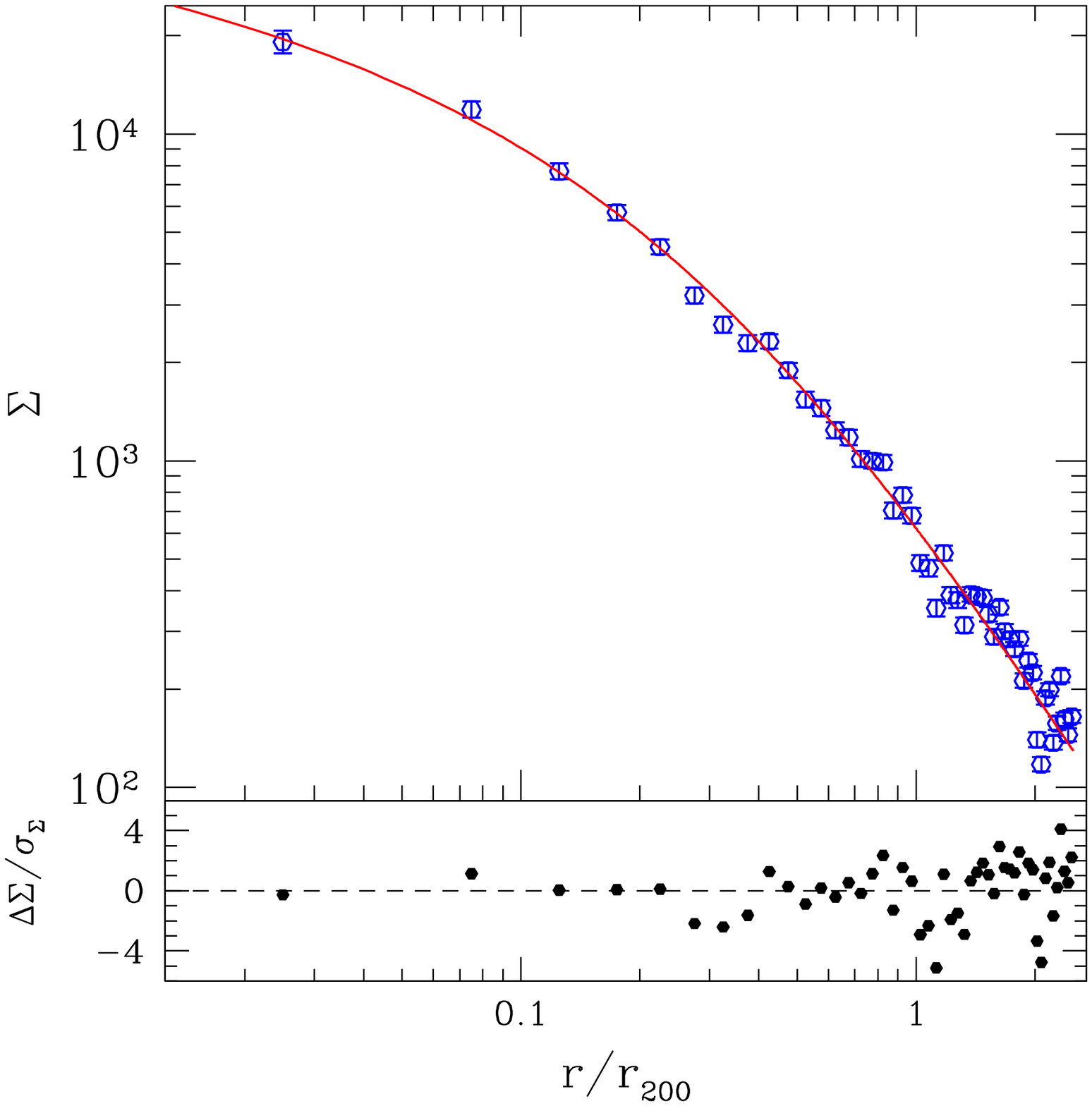}\end{minipage}
   \end{center}}
   {\myputfigure{f1.pdf}{0.0}{1.0}{-70}{-40}}
   \figcaption{\label{fig:surfden}
        Composite galaxy surface density with BCGs excluded.
        The best-fit NFW profile is
        shown. The fit is obtained by fitting the data within $r_{200}$ only.
        The figure shows the distribution of 6608 galaxies, of which 1467
        ($22\%$)
        are estimated to be background. The lower panel shows the residuals
        (differences between data and the fit) in units of standard deviation,
        which does not include the contribution from galaxy clustering.
     }
\end{inlinefigure}

The figure shows that the NFW model is a reasonable description of
the galaxy distribution out to large radius. This is consistent with the 
analyses of 14
CNOC clusters at intermediate redshifts ($z \sim 0.3$, \citealt{carlberg97b,
vandermarel00}), who found
that $c_g = 3.7$ \& 4.2, respectively. In particular, in comparison to
the galaxy surface density fit of the latter (see their equation 2),
our nearby 2MASS clusters seem to have a slightly broader distribution.
Using clusters drawn from the ENACS, \citet{adami98,adami01} find that
a profile with a core is generally preferred over a cuspy profile.
Although the distribution of fainter (e.g. $M_K>-22.5$ mag) galaxies
in our sample indeed can be well described by $\beta$-models, the
distribution of all (or brighter) galaxies is better fit by cuspy
profiles such as the NFW or a generalized-NFW profile (of the form
$n(x) \propto x^{-a} (1+x)^{a-3}$ with $1<a<2$,
e.g. \citealt{vandermarel00}). In fact, a generalized-NFW profile
with $a=1.07$, $c_g=2.71$ provides a good fit to the total galaxy distribution 
as well.

We note that the concentration parameter we obtain is the galaxy number 
weighted average over all clusters. As shown in 
\S\ref{sec:profconstraint},
the galaxy concentration in high and low mass clusters is similar, and so the
value obtained from all clusters is representative.
The concentration $c_g$ we find lies at the lower bound of measured dark matter
concentrations $c_{dm}$; there have been many attempts to study the
matter distribution within clusters, via analysis of galaxy dynamics
\citep[e.g.][$c_{dm} \sim 4-5$] {vandermarel00,biviano03,katgert04},
X--ray emission from the intracluster medium \citep[e.g.][$c_{dm} \sim
4-5$]{pratt02,lewis03}, weak/strong lensing \citep[e.g.][$c_{dm} \sim
3-8$]{clowe02,arabadjis02} or the caustic approach
\citep[e.g.][$c_{dm} \sim 4-17$]{geller99,biviano03,rines03}.  In general these
findings are in agreements with the direct numerical simulations of
the dark matter clustering \citep[e.g. NFW,][]{bullock01}.

We can also compare our result with the numerical studies in which distinct
subhaloes are resolved. Combining high-resolution numerical simulation and
semianalytic galaxy formation model, \citet{springel01b} find that galaxy
distribution in their simulated Coma-like cluster is not the same as that of the
dark matter. More recently, a study of subhalo properties within a large 
cluster ensemble shows that the subhaloes are less concentrated than the mass
\citep{delucia04}.

In the following sections, we take $c_g=3$ when converting the number
of galaxies contained in the cylindrical volume to that in the virial
sphere, necessary when we solve for LFs for individual clusters
(\S\ref{sec:lm}). On the other hand, we use $c_{dm}=5$ for conversion
between different cluster radii (e.g. $r_{500}$ to $r_{200}$).  We return to
the surface density profile in \S\ref{sec:profconstraint} for a subset
of our sample.

\subsection{Cluster Luminosity Function}
\label{sec:clf}

We stack the clusters in luminosity space to generate a composite LF
as follows.  For each cluster, we transform the apparent magnitudes of
all galaxies within the cluster virial radius $r_{200}$ to absolute
magnitudes, assuming all galaxies are at the cluster redshift. 
Galaxy magnitudes are then binned with 0.25 mag width. The faintest bin
corresponds to the bin whose faint edge is just above (brighter than)
the absolute magnitude of the 2MASS completeness limit at the cluster
redshift.  The $\log N - \log S$ relation is translated to
absolute magnitude for statistical background subtraction. We then apply the 
k-correction of the form $k(z)=-6\log (1+z)$, following \citet{kochanek01}.  
The estimated number of cluster member galaxies is divided by the cluster
volume (adjusted from a sphere to a cylinder to take the projection
effect into account, see \S\ref{sec:profile} above).

The Schechter function generally
underestimates the abundance of very bright galaxies, which is mainly due
to the presence of the BCGs \citep[e.g.][]{schechter76,goto02,christlein03}.
Their luminosity seems to be drawn from a distribution different from
the usual cluster LF.  We treat these objects separately when solving
the cluster LF for individual clusters, and also do not include them
when stacking the LFs.  We discuss the effects of the BCGs on the
light--mass relation in \S\ref{sec:lm}.

\begin{inlinefigure}
   \ifthenelse{\equal{\figtype}{EPS}}{
   \begin{center}
   \epsfxsize=8.cm
   \begin{minipage}{\epsfxsize}\epsffile{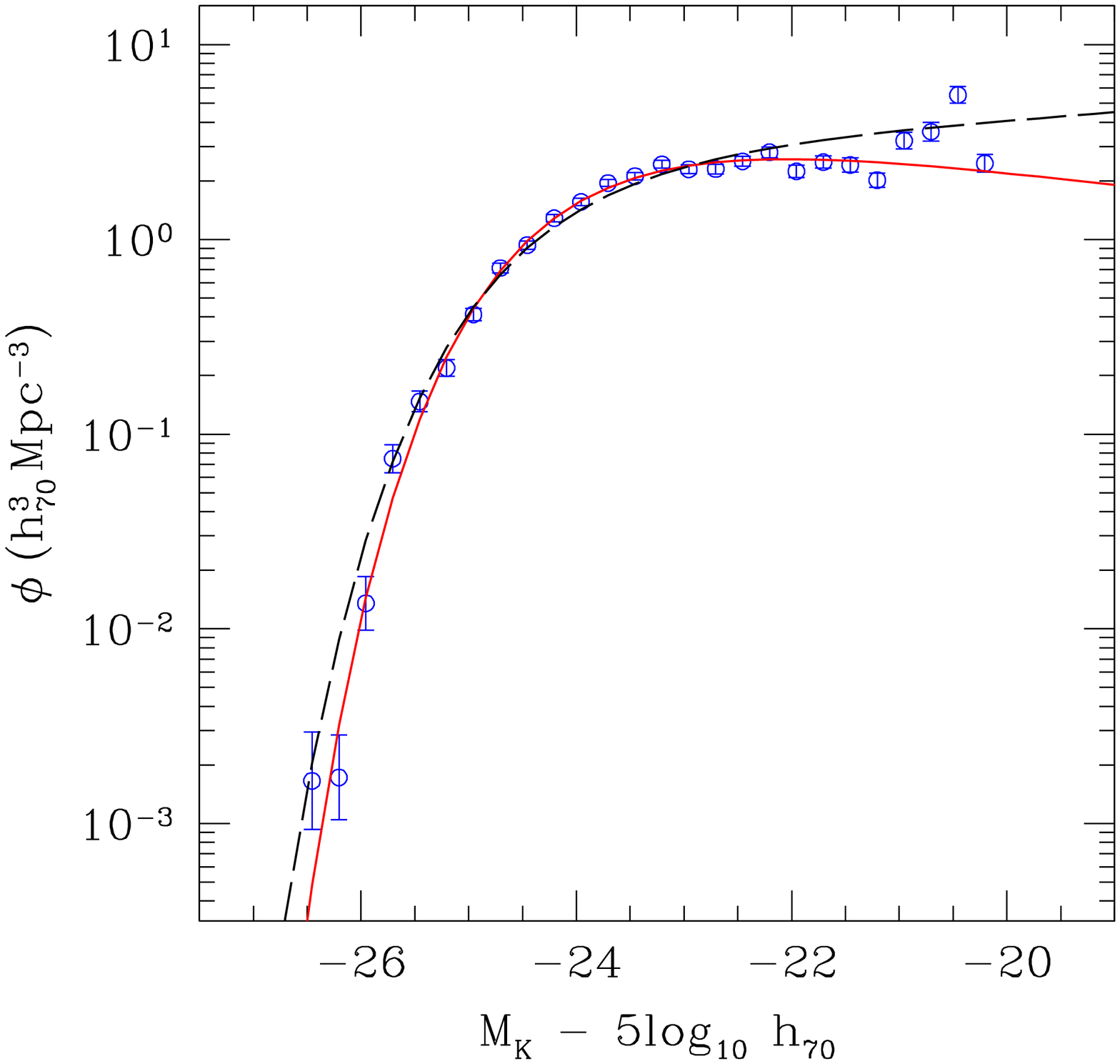}\end{minipage}
   \end{center}}
   {\myputfigure{f2.pdf}{0.0}{1.0}{-70}{-40}}
   \figcaption{\label{fig:lf}
        $K$-band composite cluster luminosity function with BCGs excluded.
        The solid line shows a Schechter function with $\alpha=-0.84$,
        $M_{K*} - 5\log_{10} h_{70} = -24.02$ and $\phi_* =4.43
        h_{70}^3$ Mpc$^{-3}$, and the dashed line shows a fit with $\alpha$
        fixed at $-1.1$, $M_{K*} - 5\log_{10} h_{70} = -24.34$ and $\phi_* =
        3.01 h_{70}^3$ Mpc$^{-3}$. The uncertainties do not include
        the contribution from galaxy clustering.
     }
\end{inlinefigure}

Figure~\ref{fig:lf} shows the composite LF, which is comprised of 5932
galaxies, of which 1256 are estimated to be background. Although the bright end 
shows the expected exponential drop, the scatter at the faint-end needs further discussion.
Statistical background subtraction is adequate when a bin contains
contributions from several clusters from different areas on the
sky. However, we notice that only a couple of clusters contribute to
the faintest bins, and this together with the growing background correction may explain the larger scatter.  Nevertheless, we expect the behavior of the 
composite LF to represent the underlying true cluster LF at $M_K
-5\log_{10} h_{70} \le -21$. In Fig~\ref{fig:lf} two fits are shown 
(obtained via chi-square fitting):
the dashed-line ($\alpha=-0.84\pm0.02$, $M_{K*} - 5\log_{10} h_{70} =
-24.02\pm 0.02$, $\phi_* =4.43\pm 0.11 h_{70}^3$ Mpc$^{-3}$) is the best-fit, 
which seems to fit most of the bins at the bright end (we notice that the small number of galaxies in the faintest bins makes their statistical weight small).
On the other hand, the solid line, which is obtained by fixing 
$\alpha$ at $-1.1$ while letting the other two parameters vary, 
may reflect the gradual rise at the faint end more faithfully
($M_{K*} = -24.34\pm 0.01$, $\phi_*=3.01\pm0.04$ Mpc$^{-3}$). Although it is 
reasonable to
assume the true value of $\alpha$ lies within these two values, in the
absence of deeper photometry (with depth to $M_K=-19$ or so) or an even larger 
sample of nearby clusters,
we will assume $\alpha=-1.1$, the same value we use in paper I, in the
following sections.  
We note that the error bars in 
Fig~\ref{fig:lf} do not include the contribution from galaxy clustering, and
therefore underestimate the true uncertainty at the faint-end; our 
choice of $\alpha=-1.1$ is therefore roughly consistent with the data. 
In $\S$\ref{sec:system} we will test the sensitivity of our results to the choice of $\alpha$.

It is important to compare our results to previous studies: \citet{balogh01a} 
use the second incremental release data from 2MASS to build a composite LF for 
clusters, based on 274 cluster galaxies with redshift measurements from the Las 
Companas Redshift Survey. Despite the large scatter due to the small number of
galaxies in their sample, they find that $M_{K*} =
-24.58 \pm 0.40$ and $\alpha=-1.3\pm 0.43$.  The agreement in $M_{K*}$
is encouraging, while the difficulty in determining $\alpha$ is
illustrated. Our $M_{K*}$ is also close to the value in more distant
clusters: $M_{K*} = -24.27 \pm 0.49$ for the $z=0.15$ bin in the
sample of \citet[][note that they assume $\alpha=-0.9$]{depropris99}. 
Using a sample of 5 clusters which span
a large range in redshift, \citet{trentham98b} estimate that 
$\alpha=-1.38 \pm 0.24$ in $K$-band and find the slope is not a strong function 
of redshift.  A study of a $z=0.3$ cluster in $K$-band finds $\alpha=-1.18$, 
and the exact value may depend on the location within the cluster
\citep{andreon01}. 
In $H$-band the existing studies show similar values
for $\alpha$ (\citealt{andreon00,tustin01}, but see \citealt{depropris98}).

In optical bands there are many systematic studies of the cluster LF. For 
example, the 2dF collaboration studies a sample of 60 clusters in $b_J$ band 
and finds that the composite LF is characterized by $\alpha=-1.28$ 
\citep{depropris03}. This study finds that there is little variation in the
LF between low and high mass clusters (see \S\ref{sec:lfconstraint}).
Using the SDSS commissioning data \citet{goto02} study a large sample of
clusters and find that the slope becomes less steep in redder bands
(see \citealt{depropris03,goto02} for comparisons between various studies).

We can also compare the cluster LF with the ``field'' LF. From 4192 galaxies
with redshift measurements, \citet{kochanek01} find that $M_{K*}
= -24.16 \pm 0.05$ and $\alpha=-1.09\pm 0.06$, using the 2MASS data. 
Recently, by combining the 2MASS and SDSS EDR photometry, \citet{bell03} find
that $\alpha=-0.77$ and $M_{K*} =-24.06$.
At slightly higher redshift ($0.1<z<0.3$), \citet{feulner03} find 
$M_{K*}=-24.56 \pm 0.24$, and $\alpha=-1.1$.  
Overall, the field and the 
cluster LF are similar, which is in agreement with the findings of previous
studies in red bands \citep[e.g.][]{trentham98b,depropris98,christlein03,andreon03}.

The cluster LF within a smaller area ($r_{500}$) is very similar to that 
obtained at $r_{200}$, with a higher 
signal-to-noise ratio with respect to the background; 
the LF is composed of 3785 galaxies, of which 549 are estimated to be 
background. The best-fit Schechter parameters are $\alpha=-0.83\pm0.02$, $M_{K*} - 
5\log_{10} h_{70} = -24.02\pm0.03$, $\phi_*=10.06\pm 0.29 h_{70}^3$ Mpc$^{-3}$; these are
almost identical to that obtained at $r_{200}$, with a larger $\phi_*$.
This larger value is expected because, by definition, the region enclosed by
$r_{500}$ is denser than that within $r_{200}$. Furthermore, the ratio
$\phi_{*,500}/\phi_{*,200}=2.27$ is intriguing: a galaxy distribution with
$c_g=c_{dm}=5$ gives $\phi_{*,500}/\phi_{*,200}=2.5$, while a $c_g=3$, 
$c_{dm}=5$ model gives
a ratio of 2.3, which is in better agreement with the above value. This is
another indication that the galaxies follow an NFW distribution, but
have a smaller concentration compared to the dark matter.

\section{Statistical Properties of Cluster Ensemble}
\label{sec:lmhod}

Using the composite LF and galaxy distribution in our data, we
calculate the total light for each cluster and examine the $L$--$M$
relation, paying particular attention to the BCGs, which we neglect
when creating the composite cluster LF and galaxy distribution.
We show that exclusion of the light contributions from the BCGs
significantly changes the $L$--$M$ relation.  Specifically, we will
see that without these BCGs, the galaxy light in clusters reflects the
galaxy number.  We discuss possible implications of this finding for
cluster formation scenarios, and we further examine
constraints on these scenarios in the next section.  
%
Table 1 contains parameters for the scaling
relations presented in this section, derived at both $r_{500}$ \&
$r_{200}$. In Table 2 we present the derived quantities (Schechter function
parameters, total light and galaxy number) for the clusters in our sample.

\subsection{Luminosity -- Mass Correlation and the Effects of the 
    Brightest Cluster Galaxies}
\label{sec:lm}

\begin{inlinefigure}
   \ifthenelse{\equal{\figtype}{EPS}}{
   \begin{center}
   \epsfxsize=8.cm
   \begin{minipage}{\epsfxsize}\epsffile{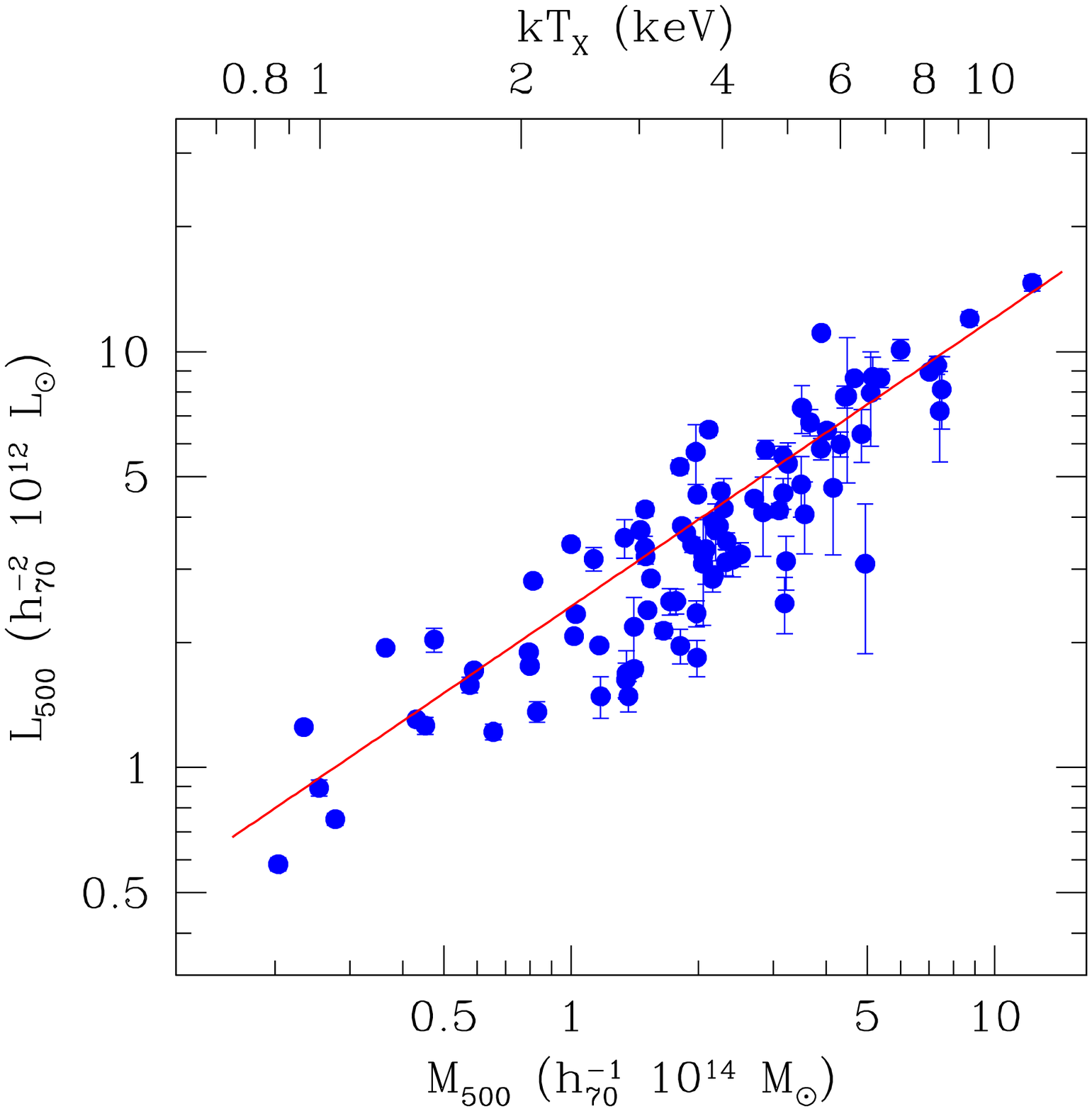}\end{minipage}
   \end{center}}
   {\myputfigure{f3.pdf}{0.0}{1.0}{-70}{-40}}
   \figcaption{\label{fig:lm}
        $K$-band luminosity--mass correlation within $r_{500}$. The best-fit
        relation has a slope of $0.69 \pm 0.04$. The scatter about the best-fit
        is $32\%$. For most of the clusters the uncertainties in light is
        smaller than the size of the points. For clarity we do not show the
        uncertainty in cluster mass (see Fig~\ref{fig:mlr}).
        At the top is the \xray temperature, from which $M_{500}$
        is estimated (see Eqn~\ref{eq:mt}).
     }
\end{inlinefigure}

We show in Fig~\ref{fig:lm} the correlation for the ensemble of 93
clusters. The systems in our sample, which span one and a half orders
of magnitude in mass, show striking regularity in their total NIR
galaxy light. The figure also shows the best-fit relation, which is
obtained by minimizing the vertical distance from the data points to
the line:
\begin{equation}
\label{eq:lm}
\frac{L_{500}}{10^{12}\,h_{70}^{-2}\,L_\odot}= 3.95\pm 0.11
    \left({M_{500}\over 2\times 10^{14}\,h_{70}^{-1}\,M_\odot}\right)^{0.69\pm 0.04},
\end{equation}
where the uncertainty is obtained from bootstrap resampling and refitting 1000
times.
The Spearman correlation coefficient is 0.86, with a probability of 
$\mathcal{O}(10^{-28})$ that such a correlation happens by chance.
As in paper I, we obtain the total light by integrating the LF of individual
clusters from a limiting luminosity corresponding to $M_{low}=-21$. Changing 
this luminosity does not affect the slope we find. However, we caution that
more massive clusters tend to be more distant (because our sample is roughly
X--ray-selected), integrating their LFs requires more extrapolation than low
mass systems. We discuss possible systematics associated with this, and the
choice of the faint-end slope of the LF in \S\ref{sec:system}.

Our result confirms the existence of the NIR $L$--$M$ relation we
found in paper I, and the slopes are consistent: based on 27 
clusters, using the second incremental release data we found $L_{500}
\propto M_{500}^{0.69\pm 0.09}$. We notice that the similarity between the 
slopes found is a coincidence: the inclusion of the k-correction, which is 
ignored in paper I, reduces the slope. Furthermore, the amplitude of the current
$L$--$M$ relation is $\sim 20\%$ smaller than that found previously. This is 
primarily due to the inclusion of the $k$-correction ($12\%$) and the different
galaxy radial profile concentration ($6\%$).

The {\it rms} scatter about the best-fit scaling relation (Eqn~\ref{eq:lm}) is 
$\sim 32\%$.
We note that because the total light depends on the cluster mass, the scatter
may well be a mix of some {\it intrinsic} scatter in the $K$-band light and that
due to the \xray $M$--$T_X$ relation.
In order to estimate the intrinsic scatter in the $K$-band light, we examine
the light contained within a fixed metric radius of $0.75$ Mpc (instead of fixed
overdensities), as this way the quantities in the two axes are independent. In 
addition to luminosity
uncertainties, we add a component $\sigma_{int}^2$ that represents the 
intrinsic scatter in the $\chi^2$ fitting. The best-estimate of $\sigma_{int}$
is such that the reduced $\chi^2=1$. We find this indicates that $24\%$ 
scatter in the $L$--$M$ relation would be intrinsic.
This impressive regularity, combined with the wide range in mass over which the 
light correlates with mass, shows the potential use of the galaxy NIR light
as a proxy for cluster binding mass (to an accuracy $\lesssim 43\%$), 
which will be very useful in optical/NIR cluster surveys. We further 
investigate this aspect in \S\ref{sec:summary}.

\begin{inlinefigure}
   \ifthenelse{\equal{\figtype}{EPS}}{
   \begin{center}
   \epsfxsize=8.cm
   \begin{minipage}{\epsfxsize}\epsffile{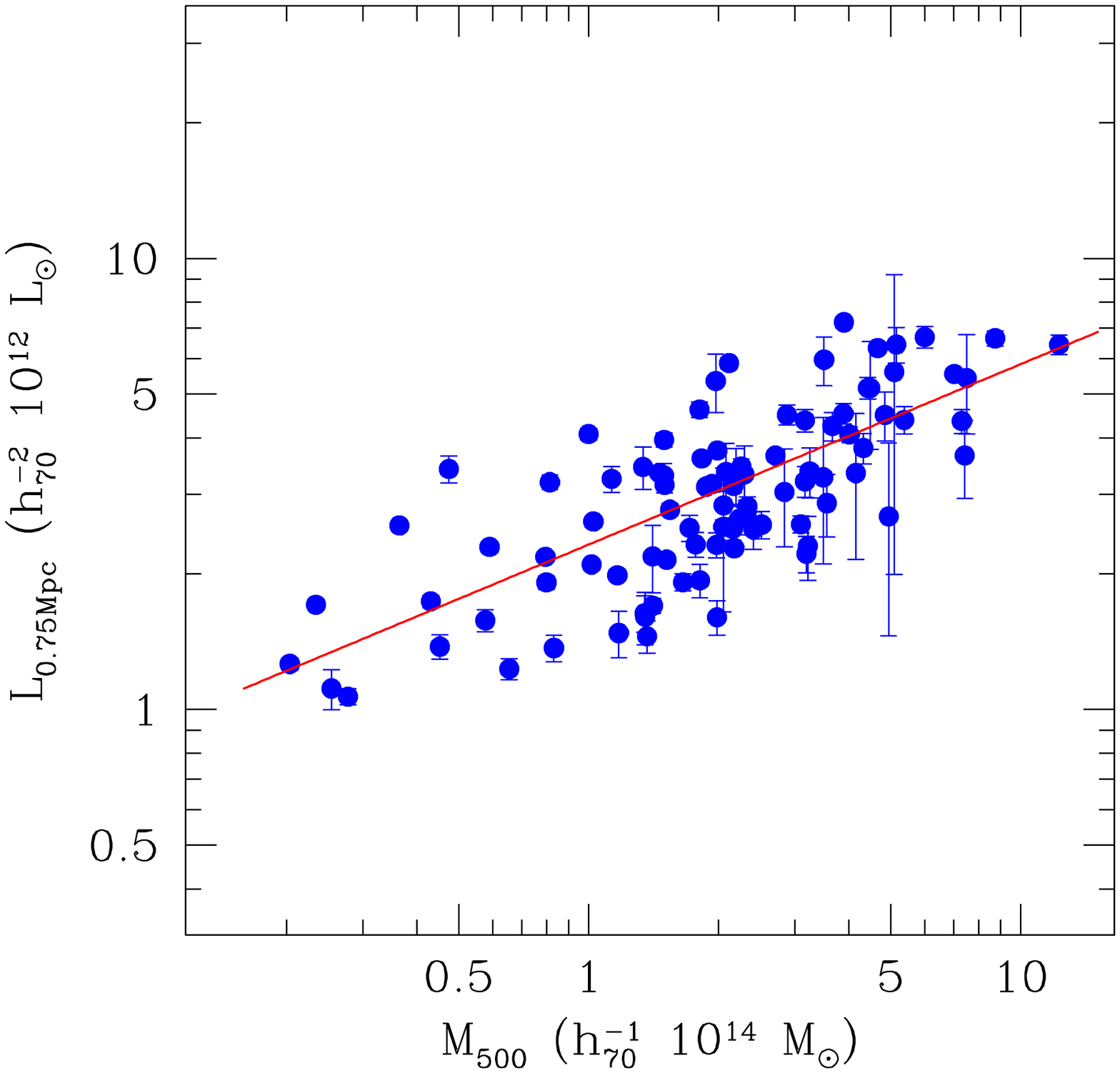}\end{minipage}
   \end{center}}
   {\myputfigure{f4.pdf}{0.0}{1.0}{-70}{-40}}
   \figcaption{\label{fig:mpc}
        Total $K$-band galaxy light within a fixed metric radius of 0.75 Mpc.
        The correlation indicates that 24\% scatter in the $L$--$M$ relations
        (e.g. evaluated at fixed overdensities) is intrinsic.
     }
\end{inlinefigure}

The $L$--$M$ relation measured within $r_{200}$ is
\begin{equation}
\label{eq:lm200}
\frac{L_{200}}{10^{12}\,h_{70}^{-2}\,L_\odot}= 5.64\pm 0.16
    \left({M_{200}\over 2.7\times
    10^{14}\,h_{70}^{-1}\,M_\odot}\right)^{0.72\pm 0.04},
\end{equation}
where the normalization mass is the $M_{200}$ corresponding to that
used in Eqn~ \ref{eq:lm}. In obtaining this, we solve for individual
cluster LFs at $r_{200}$.  The individual $L_*$s are not very different
from those solved at $r_{500}$; the mean of the ratio of the two
quantities for all the clusters is
$\overline{L_{*,500}/L_{*,200}}=1.07$.  However, as we note in
\S\ref{sec:clf}, the characteristic density is very different. We have
$\overline{\phi_{*,500}/ \phi_{*,200}}=2.22$, which is very close to
the ratio obtained from the fits to the composite LFs.
Furthermore, the ratio of the mass-to-light ratios evaluated at the
two radii is $\overline{\Upsilon_{500}/\Upsilon_{200}}=1.04$. This
means that the cluster mass-to-light profile is only slightly
decreased toward larger radii, which is consistent with some previous
findings (e.g. \citealt{rines01b,katgert04}; see \citealt{biviano03}
for a review).  An equivalent way of stating this is that light increases only 
a bit faster with radius than the mass does (our results imply that
$L_{200}/L_{500}=1.43$, compared to $M_{200}/M_{500}=1.38$ for an NFW
profile with $c_{dm}=5$).

\begin{inlinefigure}
   \ifthenelse{\equal{\figtype}{EPS}}{
   \begin{center}
   \epsfxsize=8.cm
   \begin{minipage}{\epsfxsize}\epsffile{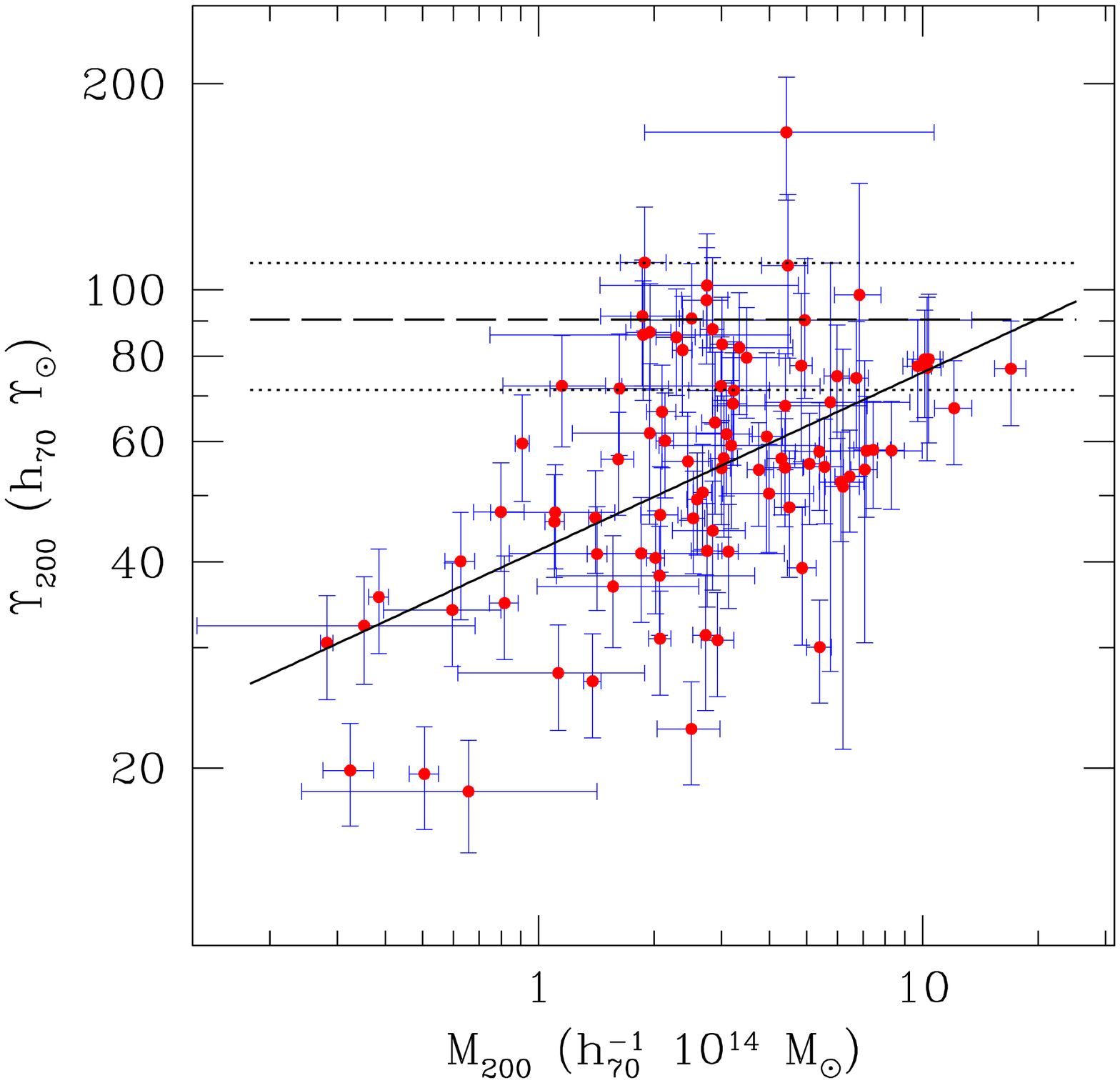}\end{minipage}
   \end{center}}
   {\myputfigure{f5.pdf}{0.0}{1.0}{-70}{-40}}
   \figcaption{\label{fig:mlr}
        $K$-band \mlr within $r_{200}$. The solid line shows the best-fit
        relation, which has slope $0.26 \pm 0.04$. The dashed line shows the
        universal \mlre, derived using the \Om measured by {\it WMAP} and the
        luminosity density estimated by \citet{bell03}. The dotted lines show
        the uncertainties in $\Upsilon_{univ}$.
     }
\end{inlinefigure}

An immediate implication of the $L$--$M$ relations found here is that the 
cluster mass--to--light ratio is an increasing function of mass. 
Figure~\ref{fig:mlr} shows the \mlr calculated at $r_{200}$, $\Upsilon_{200}$. 
The best-fit relation is almost identical to
that inferred from Eqn~\ref{eq:lm200}. The uncertainties in $\Upsilon_{200}$
is obtained by summing in quadrature the contributions from the cluster radius, 
total light, the dark matter
and galaxy concentrations. Our finding of an increasing \mlr is consistent with 
other studies (in $K$-band: paper I; in optical: 
\citealt{girardi02,bahcall02}, and in group scales: \citealt{pisani03,tully03}).
It is interesting to compare the cluster \mlr with that of the universe
$\Upsilon_{univ} = \Omega_M \rho_c/\bar{j} = 90\pm 19 h_{70} 
\Upsilon_\odot$ 
(shown as the dashed line in the figure), where we take the measurement of
\Om from {\it WMAP} \citep{bennett03}, and use the mean luminosity density 
$\bar{j}$ measured by \citet{bell03}. 
The mass dependence of cluster \mlr makes it less secure as an indicator
of the matter density parameter, unless the relative distribution of light and
matter is taken into account \citep{ostriker03}.

The existence of the $L$--$M$ relation (or an increasing \mlre) is intriguing 
in two ways. First, it indicates that the star or galaxy formation
process in clusters and groups is regular. It is thus relatively
easy to predict the stellar mass in a cluster, once a mean stellar \mlr
is known. (Please see \S 4.1 in paper I for further discussion.)
Second, the
slope of the scaling deserves more attention; because the slope is
less than unity, high mass clusters seem to produce light (stars) less
efficiently than low mass clusters. In paper I we attribute this
to a possible change in star formation efficiency across this mass
range. Although this is one viable cause, we will seek other possible 
explanations within the context of the HOD. For now we remind the reader that we
are only accounting for light within galaxies.

The brightest cluster galaxies are among the most luminous objects in
the universe. Their contribution to the cluster luminosity is not
negligible \citep[e.g.][]{oemler76,gonzalez00}. We have identified the BCGs in
the clusters in our sample and estimate their
luminosity. Interestingly we find that, although the distribution of
the BCG luminosity $L_{bcg}$ is relatively narrow (in terms of
$K$-band magnitude, the mean is $M_K = -26.18 \pm 0.05$ mag, with a
0.45 mag scatter), there exists a tight correlation between $L_{bcg}$
and the total luminosity $L_{500}$ (which is the sum of the luminosity
from all galaxies).  These results will be presented elsewhere (paper III).
The most important implication from this
investigation is that the BCG contribution to the total light is a
crucial factor in shaping the $L$--$M$ correlation.  Because a
larger proportion of the light in group scale systems comes from the
BCG, exclusion of $L_{bcg}$ steepens the slope of the light--mass
relation.  Denoting the total light from all galaxies but the BCG as
$L_{500}^{s} = L_{500}-L_{bcg}$, it is found that $L_{500}^{s} \propto
M_{500}^{0.82}$ -- a $3\sigma$ change in slope!

The BCGs are the main cause of a flatter $L$-$M$ relation. 
We turn to the halo occupation number now to further examine the origin of the 
$L^{s}$--$M$ relation.

\subsection{Halo Occupation Number}
\label{sec:hod}

We examine the halo occupation number using the galaxy LF within each cluster.  Once
the Schechter parameters $\phi_*$ \& $L_*$ are determined for each cluster
(without the BCG),  the total number of galaxies brighter than a luminosity threshold $L_{low}$ 
is simply $N = 1+N^s$, where the first term denotes the BCG count, and 
$N^s = V \phi_* \int_{y_{low}}^{\infty} y^\alpha {\rm e}^{-y} d y$,
where $y_{low} = L_{low}/L_*$ and $V$ is the cluster volume.

Excluding the BCGs results in a $L$--$M$ relation at $r_{500}$ of
slope $0.82$ (c.f. Eqn~\ref{eq:lm}, Table 1). This demonstrates the importance
of light from BCGs to the total luminosity budget. However, unlike the
total luminosity, the mean number of galaxies should be much less
sensitive to the absence of BCGs, since they only stand for one count
per cluster.  Including the BCGs, we find that $N_{500} \propto
M_{500}^{0.82\pm 0.04}$; without the BCGs, the resulting halo
occupation number is (Fig~\ref{fig:hod})
\begin{equation}
\label{eq:hod}
N_{500}^s (M_K\le -21) = 56 \pm 2
    \left({M_{500}\over 2\times 10^{14}\,h_{70}^{-1}\,M_\odot}\right)^{0.84\pm 0.04},
\end{equation}
and indeed only small change in slope is found. The superscript emphasizes this
number does not count the BCGs. We note that the {\it rms} scatter about the
best-fit line is $35\%$, slightly larger than that of the light--mass relation.
The halo occupation number, as well as other derived LF parameters, for each
cluster, is listed in Table 2.

Interestingly, the slope of the $N^s$--$M$ relation is essentially the
same as the BCG-free $L^s$--$M$ relation. This implies that, except for
the BCGs, the sum of galaxy luminosities scales as total galaxy
number in clusters (i.e. the shape parameter $L_*$ of the luminosity function 
does not change with cluster mass). This is true also for the pair of relations
evaluated at $r_{200}$: $L_{200}^{s} \propto M_{200}^{0.82\pm 0.04}$, and
$N_{200}^{s} \propto M_{200}^{0.87\pm 0.04}$ (Table 1).

The bottom panel of Fig~\ref{fig:hod} shows $\alpha_p$, a measure of
the spread in $P(N|M)$, defined as the ratio of the second moment to
the first moment of the occupation number
\[
\alpha_p^2 \equiv \left< N(N-1) \right>/ \left< N \right>^2
\]
\citep{berlind03,kravtsov03}.  In the case that the distribution is
Poisson, $\alpha_p = 1$. 
We calculate $\alpha_p$ by averaging the occupation numbers in 7 mass bins, 
each of which has a roughly equal number of clusters.

\begin{inlinefigure}
   \ifthenelse{\equal{\figtype}{EPS}}{
   \begin{center}
   \epsfxsize=8.cm
   \begin{minipage}{\epsfxsize}\epsffile{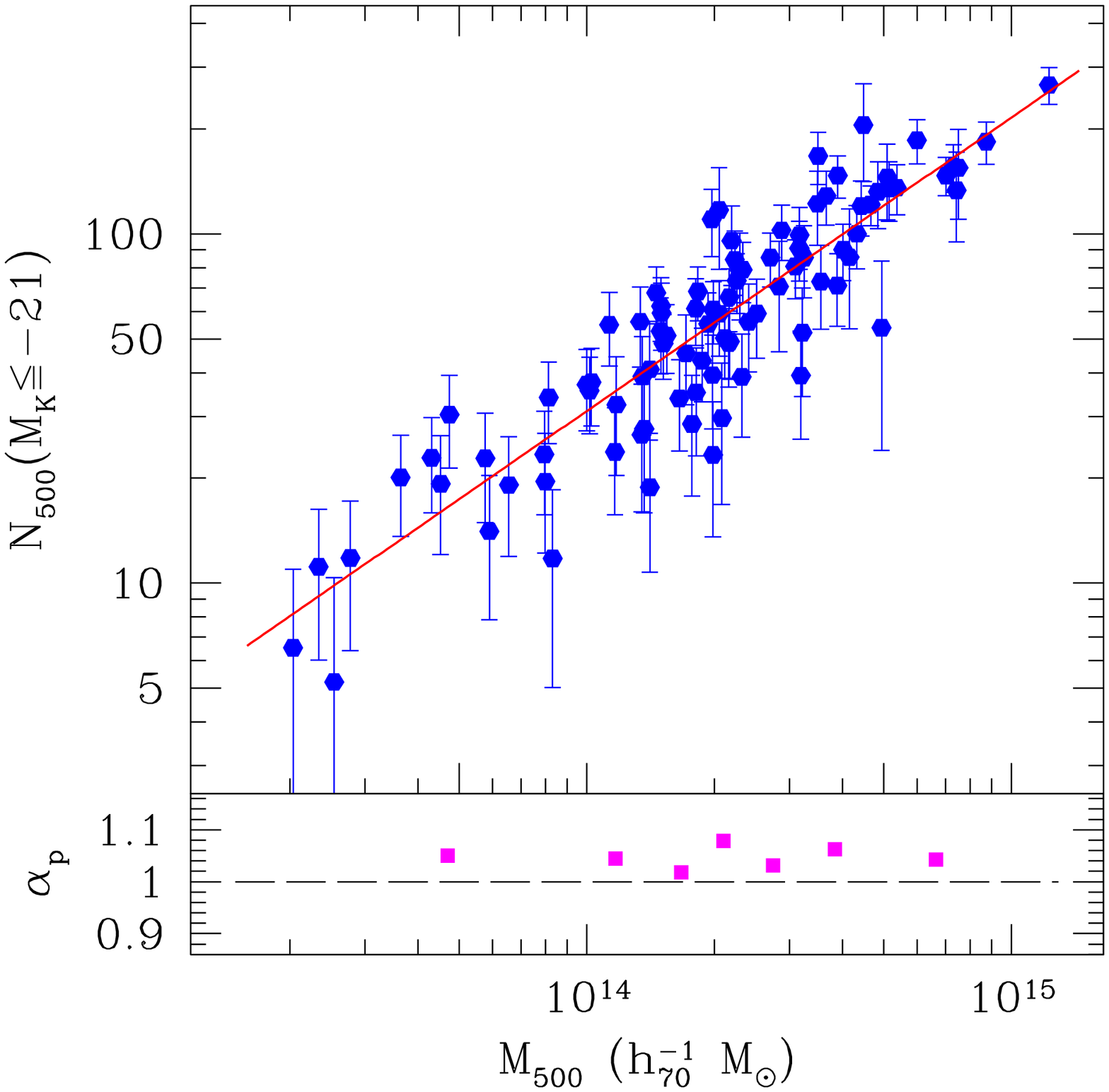}\end{minipage}
   \end{center}}
   {\myputfigure{f6.pdf}{0.0}{1.0}{-70}{-40}}
   \figcaption{\label{fig:hod}
        {\it Top}: Number of galaxies as a function of cluster mass. The
        best-fit relation gives $N \propto M^{0.84\pm0.04}$.
        {\it Bottom}:
        deviation of the halo occupation number from a Poisson distribution.
        $\alpha_p = 1$ for Poisson, while a narrower (broader) distribution has
        $\alpha_p<1$($>1$).
     }
\end{inlinefigure}

There are several observationally determined results for the halo
occupation number. Most of these come from matching the observed
galaxy two-point correlation function or power spectrum
\citep[e.g.][]{seljak00,peacock00, yang03, zehavi03,
magliocchetti03}. Generally speaking, this is done by assuming a form
of the HOD, adjusting the parameters until the prediction from the
halo model matches the observed galaxy clustering. There are other
groups who use different approach to study the HOD.
For example, by studying the velocity dispersions of groups in the Updated
Zwicky Catalog, \citet{pisani03} find that $N\propto M^{0.70\pm0.04}$.
Using the LF from the Nearby Optical Galaxy
sample, \citet{marinoni02} obtain the HOD for objects with a wide
range in mass. 
An advantage of using the LF approach to determine the properties of the HOD
is that we can measure any moment of the HOD, unlike the approach relying on
the measurements of galaxy clustering, which is sensitive to the first and
second moments only.

Another approach similar to our use of the NIR $N$--$M$ relation is presented in
\citet[][hereafter K03]{kochanek03}.  They identify clusters in the 2MASS
all-sky data using a matched filter cluster finding algorithm. They
also calculate the $N$--$M$ based on a sample of 84 clusters with \xray
temperatures and find that $N_{666} \propto T_X^{2.09\pm 0.17}$, where
$N_{666}$ is the number of galaxies brighter than $M_*=-24.16$ within
$r_{200}$. With the \xray mass--temperature relation $M_{500}
\propto T_X^{1.58}$ adopted in our work, their relation goes as $N_{666} \propto
M_{500}^{1.32\pm 0.11}$, which is inconsistent with our result. K03 examine the
$N$--$M$ relation using several different mass estimators, and the average slope
is $\gamma = 1.10 \pm 0.09$, closer to our results.  

We have considered possible explanations for these differences.  A way to bias the slope of the $N$--$M$ relation is to have a mass-dependent systematic in the analysis. In both studies a constant
faint-end slope $\alpha$ is assumed (although the values differ slightly).   As shown below in $\S$\ref{sec:lfconstraint}, there is no evidence for variations in $\alpha$ with mass to a depth of $M_K=-21$.  
They report the number of galaxies brighter than $M_*$, which they assume is 
fixed, whereas we report the number of galaxies brighter than $M_{low}=-21$, a 
reasonable effective absolute magnitude limit for our cluster sample.  We fit 
for $M_*$ independently in each cluster, but there is no trend for varying
$M_*$ with mass, so this cannot explain the difference.   
As discussed in paper I, we believe that the most important difference between the K03 analysis and ours is the cluster selection and mass estimation.  
Specifically, we use the peak in X-ray emission as the cluster center and the 
emission-weighted mean temperature as the mass estimate.  Thus, we examine 
galaxy properties within a region that is defined using X-ray observables.  K03 
find their clusters and estimate galaxy number and light using only NIR 
observables.  
NIR observables are less accurate mass estimators (see Figure~\ref{fig:lm}), introducing additional scatter in the observed galaxy number at a fixed mass.  Note that the scatter about their $N$--$M$ relation derived for systems with \xray
temperature (their eq. 35) 
is approximately 2.5 times larger than the scatter about ours.  
Because of the steepness of the mass function, this scatter will tend to move more low mass systems to higher estimated mass than high mass systems to lower estimated mass;  this will generically lead to a systematic bias in the slopes of scaling relations.  
In estimating cluster light or galaxy number K03 use a galaxy model with fixed metric radius, and then attempt 
to bootstrap to the correct light or number using the properties of the detected clusters.  
It is well known that a matched filter approach works best when the target and 
filter are most similar, and so this approach could easily be a source of 
mass-dependent systematics in their analysis;  K03 work hard to tune their 
method using N-body simulations to remove these systematics.  
We now know this approach is likely to be flawed, because our Figure~\ref{fig:surfden} shows that the galaxies are distributed in a way consistent with an NFW model with concentration $c\sim3$, 
which is less concentrated than the observationally estimated dark matter 
profiles in clusters (see discussion in $\S$\ref{sec:profile}).  Thus, 
accurately tuning a cluster finding algorithm with simulations requires a 
galaxy population that has different properties than the dark matter.

Most of the above studies and our results show that at large halo
masses, $N^s \propto M^\gamma$, where $\gamma \lesssim 1$. From the
theoretical view point, this is expected. For example, in the standard
paradigm of semianalytic models galaxies are assumed to form in dark
haloes of mass above a threshold, below which there is not a
sufficient amount of gas to cool to initiate galaxy formation. At the
low mass end of the halo population, there are 1 or 0 galaxies per halo, on
average. As the structure grows hierarchically, the small haloes merge
and galaxies merge (to the central galaxies of the newly formed halos,
within a dynamic friction time). Although gas in the larger mass
haloes still can cool and form galaxies, the efficiency of gas cooling
is a steeply decreasing function of halo mass \citep[e.g.][]{cole00}. 
In this picture,
therefore, the mass naturally increases faster than the number of galaxies.  

Recent high resolution simulations have confirmed these expectations
\citep[e.g.][]{berlind03,kravtsov03}. In particular, the latter, a
pure $N$-body simulation, finds that $\gamma \sim 1$ for subhaloes of
different abundances, as well as at different cosmic epochs. By
matching the number density of their subhaloes to that of galaxies at
$M_r \le -18$ (roughly corresponding to $M_K = -21.3$, using the field
LF from \citealt{kochanek01}), they find $\gamma = 0.92 \pm 0.03$, in
good agreement with our result at $r_{200}$.

A slope $\gamma < 1$ indicates that more massive clusters have fewer
galaxies per unit mass, compared to their
lower mass counterparts. A naive explanation is that clusters ``lose''
galaxies as they merge toward the high mass end. To be more specific,
$\gamma \sim 0.84$ implies that a $10^{15} M_\odot$ cluster has $\sim
32\%$ fewer galaxies than the sum of ten $10^{14} M_\odot$
clusters. (Of course, a $10^{15} M_\odot$ cluster need not be a direct
merger remnant of ten $10^{14} M_\odot$ clusters.)  A natural question
is then, how do the galaxies disappear?  An equivalent way of thinking
about this is to ask: can we make a present day $10^{15} M_\odot$
cluster by merging less massive clusters or groups we see today? How
will the merger process alter the population of galaxies within their
progenitor haloes?  We continue in the next section to study the
implications of the $L$--$M$ and $N$--$M$ relations.

\begin{table*}[htb]
\begin{center}
\caption{$L-M$ Correlation \& Halo Occupation Number}
\begin{tabular}{rcccccc}
\tableline \tableline
 & \multicolumn{2}{c}{$L_{\Delta,12}=A M_{\Delta,14}^B$} & \multicolumn{2}{c}{$N_{\Delta} =C M_{\Delta,14}^D$} &   \multicolumn{2}{c}{$\sqrt{N_\Delta(N_\Delta-1)} =E M_{\Delta,14}^F$}\\
\cline{2-3} \cline{4-5} \cline{6-7}
Method & $A$ & $B$ & $C$ & $D$ & $E$ & $F$ \\
\tableline
$\Delta=500$: with BCG & $2.44\pm 0.11$ & $0.69\pm 0.04$ & $32\pm 2$ & $0.82\pm 0.04$ & $31.70^{+1.52}_{-1.45}$ & $0.83 \pm 0.04$ \\
no BCG & $1.85 \pm 0.11$ & $0.82\pm 0.05$ & $31\pm 2$ & $0.84\pm 0.04$ & $30.39^{+1.50}_{-1.43}$ & $0.85 \pm 0.04$ \\
\tableline
$\Delta=200$: with BCG & $2.75\pm 0.16$ & $0.72\pm 0.04$ & $37\pm 3$ & $0.85\pm 0.04$ & $36.79^{+1.98}_{-1.88}$ & $0.86 \pm 0.04$ \\
no BCG & $2.19\pm 0.15$ & $0.82\pm 0.04$ & $36\pm 3$ & $0.87\pm 0.04$ & $35.60^{+1.95}_{-1.85}$ & $0.88 \pm 0.04$ \\
\tableline
\end{tabular}
\tablecomments{ $L_{\Delta,12} \equiv L_\Delta/(10^{12} h_{70}^{-2} L_\odot)$;
    $M_{\Delta,14} \equiv M_\Delta/(10^{14} h_{70}^{-1} M_\odot)$.
  }
\end{center}
\vskip-25pt
\end{table*}

\section{Galaxy Population Transformation}
\label{sec:further}

Within the hierarchical structure formation scenario, high mass clusters are
made of low mass clusters through the processes of merging and accretion. 
If nothing
destroys galaxies in the merging process (which is unlikely) or
through some other process that does not have a dependence on cluster
mass, we would expect $N \propto M$, and other properties concerning
cluster baryons such as the stellar mass fraction and the cluster \mlr
to remain the same in high and low mass clusters.  However, observations show 
that this is not the case (e.g. paper I, \citealt{sanderson03}).
Identifying physical processes whose efficiency varies with cluster mass is therefore an important step toward understanding the meaning
of the $L$--$M$ and $N$--$M$ relations.

We first list a few candidate processes that may change the galaxy populations 
as clusters merge and grow in mass (e.g. \citealt{treu03}): 
tidal stripping, ``galaxy 
harassment'' (cumulative tidal disruption due to impulsive encounters between
galaxies or cluster substructures, \citealt{moore98}), dynamical friction,
galaxy mergers, ram pressure stripping, variable star formation
efficiency, and the ageing of stellar populations.  These mechanisms
may not be independent of each other, but their action can transform
galaxies' internal properties, making them fainter, or changing their
spatial distribution.  In the following sections we attempt to use the observational constraints from our 2MASS cluster study to gauge the relative importance of the mechanisms listed above.

\subsection{Constraints from Luminosity Function}
\label{sec:lfconstraint}

\begin{inlinefigure}
   \ifthenelse{\equal{\figtype}{EPS}}{
   \begin{center}
   \epsfxsize=8.cm
   \begin{minipage}{\epsfxsize}\epsffile{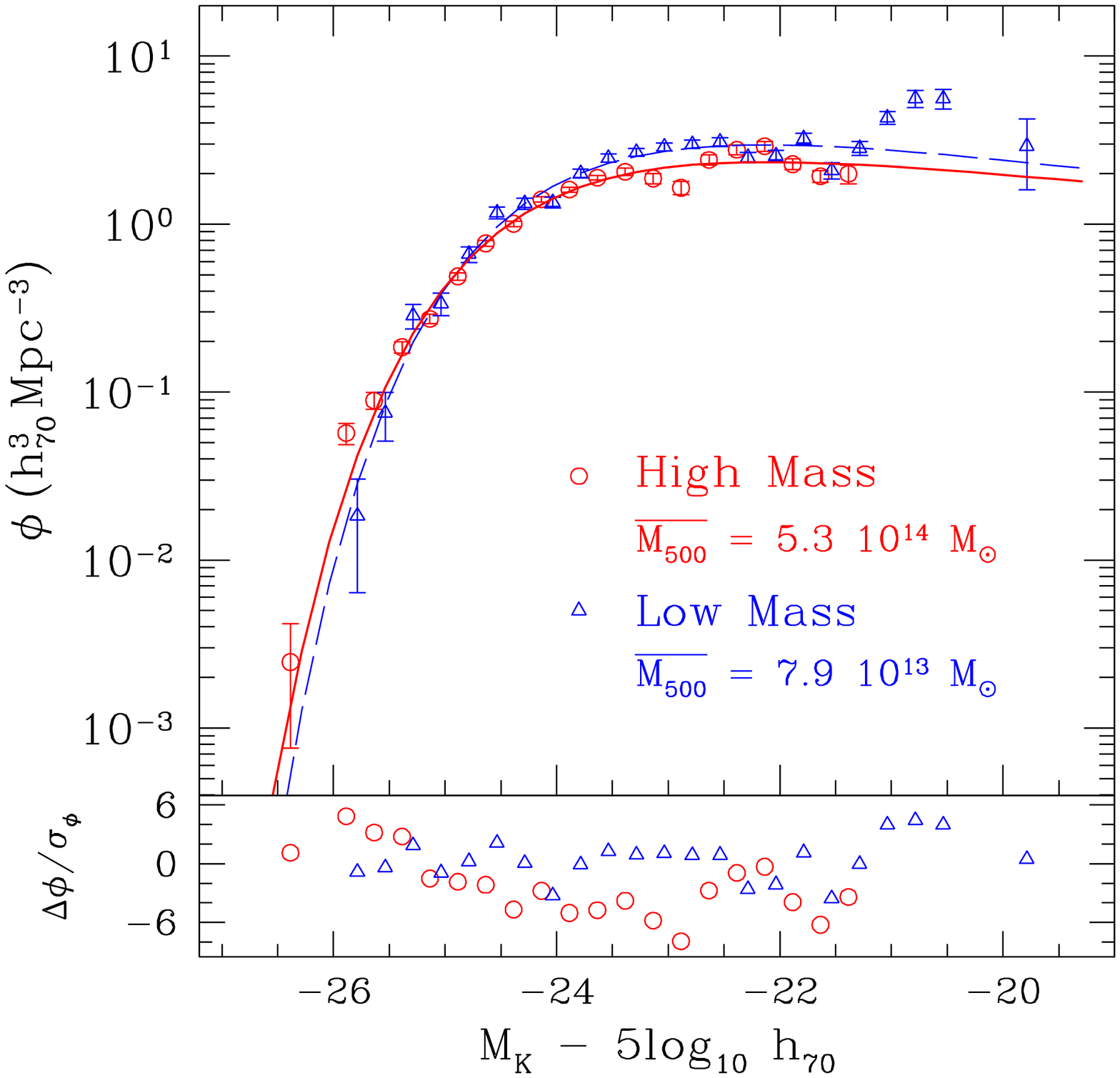}\end{minipage}
   \end{center}}
   {\myputfigure{f7.pdf}{0.0}{1.0}{-70}{-40}}
   \figcaption{\label{fig:cflf}
        Comparison of LFs for most and least massive 25 clusters. The best-fit
        Schechter functions are shown (solid line: high mass cluster LF; dashed
        line: low mass cluster LF). The faint-end slopes are similar
        (down to the completeness limit $M_{K,low}=-21$ of the high mass
        clsuter LF).
        The lower panel shows the difference between data points with the
        best-fit LF for low mass clusters. Low mass clusters show
        deficit of extremely bright galaxies, but have higher density for
        galaxies of
        moderate luminosities ($M\gtrsim M_*$) compared to high mass clsuters.
     }
\end{inlinefigure}

In Fig~\ref{fig:cflf} we show the composite LFs for the 25 most and
least massive clusters. The mean (median) masses $M_{500}$ for the two 
subsamples are $5.3 (4.7) \times 10^{14} M_\odot$ and $7.9 (8.0) \times 10^{13} 
M_\odot$, respectively.  Because the high mass clusters in our sample
tend to be at higher redshifts, their LF is only sampled down to $M_K
\approx -21$, while the LF of the low mass clusters is probed about one 
magnitude deeper.  For meaningful comparison, we only consider down to $M_K = 
-21$.

The best-fit to the high mass cluster LF (hereafter HMLF) is $\alpha=-0.84\pm 
0.03$, $M_* =-24.10\pm 0.04$ \& $\phi_*=4.00\pm 0.16$ Mpc$^{-3}$, while that for
the low mass cluster LF (hereafter LMLF) is $\alpha=-0.81 \pm 0.04$, 
$M_* =-23.94\pm 0.06$ \& $\phi_*=5.34\pm 0.33$ Mpc$^{-3}$. We note that in both 
LFs the BCGs are excluded.  From the figure it is seen that
the low mass clusters lack extremely bright galaxies (therefore the fainter
$M_*$) but are more abundant in faint galaxies (thus the larger $\phi_*$). 
To make this clearer
we plot in the lower panel of the figure the deviation of the low and high mass
cluster LFs from the best-fit LMLF, in units of standard
deviation.  At $M_K \gtrsim -23.5$ the LMLF points are higher than the 
HMLF, and at $M_K \lesssim -26$ there are no galaxies in low mass clusters.

The mechanisms that build the cluster halo occupation distribution must also be
responsible for transformation of the LFs. One naive way to convert the LMLF to
the HMLF is to reduce the abundance of galaxies fainter than $M_*$ in low mass
clusters, and make more bright galaxies that are of comparable luminosity as the
BCGs in low mass clusters. Tidal stripping, ram pressure stripping, galaxy
harassment, differences in star formation efficiency, and stellar ageing, can be
responsible for the first step, while dynamical friction and mergers between 
galaxies may act to complete the second step. We examine their effects in more
detail below.

Because of the tidal field of the host halo, subhaloes orbiting within a cluster
will lose their mass continuously \citep[e.g.][]{merritt84,klypin99,hayashi03,
delucia04}.  Utilizing the expressions in \citet{klypin99}, we estimate the 
average fractional mass loss of a galactic halo of mass $m$ within a cluster of 
mass $M$ to be 
\[
{\Delta m\over m}(m,M) = \int_0^{r_v} {\Delta m\over m} (m,M,r) \mathcal{L}(r) 
4\pi r^2 dr,
\]
where $r$ is the orbit radius, $r_v$ is the cluster virial radius, and 
$\mathcal{L}(r)$, the likelihood of galaxy distribution within the cluster, is
taken to be an NFW profile with $c_g = 3$. For $10^{11} \le m < 10^{13} 
M_\odot$, we find that $\Delta m(m,M)/m \sim 0.35-0.4$ for $M=10^{14}
M_\odot$. For larger cluster halo mass $M$, the mass loss is somewhat smaller, which is due to
the smaller concentration of the more massive halo.
These estimates may well be lower limits, as the simulations suggest 
\citep{hayashi03}. Estimating the amount of light that is stripped off is more problematic, because it involves knowledge of the relative distribution of light and mass
within subhaloes. Nevertheless, this simple exercise shows tidal stripping
may effectively make galaxies dimmer-- if not completely destroy them-- particularly in less
massive clusters.

Ram pressure stripping has been proposed to explain the possible transformation
between normal spiral galaxies and S0s in intermediate redshift clusters
\citep{poggianti99}. Numerical studies show that, with a time scale of $\sim
10^7$ yr, ram pressure stripping can deplete $\sim 50\%$ of gas from a disk
galaxy \citep{abadi99}. However, this mechanism is most
efficient in the central regions of the clusters, where the intracluster gas is
densiest; it is therefore important to check the spatial distribution of 
galaxies in order to assess the contribution of ram pressure stripping.  Regardless, the changing of galaxy type is likely not relevant in explaining the systematic variation in the luminosity function as we move from low mass to higher mass clusters.

Tidal interactions during fast encounters between galaxies or between a galaxy
and a local substructure (galaxy harassment) is an effective way of transforming
galaxies \citep{moore98}.
As numerical studies show \citep{gnedin03b}, tides are not
necessarily strongest during core penetration. In fact, most of the strong
tidal forces appear near encounters with local substructure
(e.g. within infalling groups). As clusters continue to grow by accreting
haloes, more and more galaxies get transformed (spirals become S0s,
low surface brightness galaxies and dwarf spheroidals get destroyed and produce 
diffuse intracluster light, while ellipticals, which may be in place well before
clusters form, are less affected). It is found that low surface brightness
galaxies with luminosity near $L_*$ are destroyed most efficiently 
\citep{gnedin03b}.   However, they probably only constitute a small proportion of
the galaxy population, and galaxies of this sort are unlikely to appear in our 2MASS sample \citep{bell03}; therefore, their destruction could not account for the
slope of the HOD in our analysis.

Another way of altering the halo occupation number above some limiting absolute magnitude is to alter the luminosity of galaxies, moving them into or out of the
sample.  Varying star formation efficiency across cluster mass scales acts in the right direction to reduce the number of galaxies in higher mass clusters \citep[e.g.][]{cole00,springel03}.   As for 
stellar ageing, studies indicate that stellar ages in high and low mass clusters do
not differ much \citep[e.g.][]{stanford98}.  Using ageing effects to explain our $K$-band results is especially difficult, because of the reduced sensitivity to young stars.  The NIR stellar \mlr varies only a factor of 2 over a Hubble time \citep{madau98}. 
We emphasize, though, that neither of these physical processes are at work in the numerical study of the halo occupation distribution where a similarly, flat 
$N$--$M$ relation is seen \citep{kravtsov03}.  
Thus it is clear that the galaxy transformation we observe can be explained without resorting to these processes.

As for the bright end of the LF, the appearance of bright galaxies in high mass 
clusters may be the result of merging; some of the bright galaxies may be merger
remnants occurring in lower mass haloes, and some may be the BCGs in less massive
clusters that have recently merged with the main cluster.  Because tidal stripping is very effective, producing very bright galaxies from what were previously 
BCGs in subclusters may require merging with other less luminous galaxies during cluster infall and subsequent virialization. 

In summary, the transformation of the LFs may well involve several mechanisms.  In the next section, we try to further differentiate these processes by 
inspection of the galaxy spatial distribution.

\subsection{Constraints from Galaxy Distribution}
\label{sec:profconstraint}

One way of flattening the slope of the $N$--$M$ relation would be if galaxy 
distributions in high mass clusters are more extended (with respect to the 
virial radius) than in low mass clusters.  We check this possibility by 
examining the galaxy distributions in low and high mass clusters.

\begin{inlinefigure}
   \ifthenelse{\equal{\figtype}{EPS}}{
   \begin{center}
   \epsfxsize=8.cm
   \begin{minipage}{\epsfxsize}\epsffile{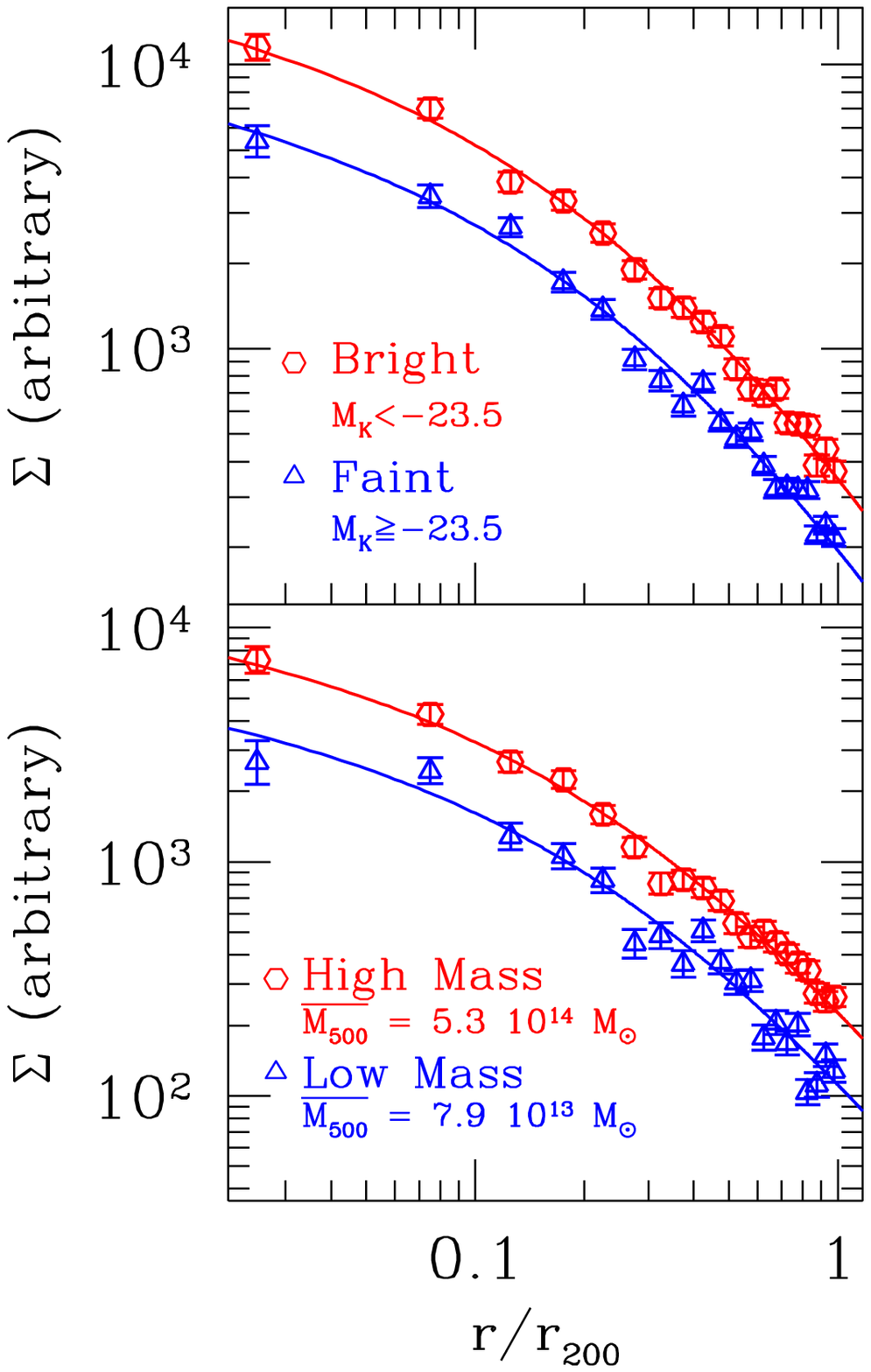}\end{minipage}
   \end{center}}
   {\myputfigure{f8.pdf}{0.0}{1.0}{-70}{-40}}
   \figcaption{\label{fig:cfprof}
        {\it Top:} Comparison of profiles for galaxies brighter and fainter than
        $M_K=-23.5$ in all clusters. The best-fit profiles have
        $c_g = 3.0\pm0.3$ (bright) \& $2.8\pm0.3$ (faint).
        {\it Bottom:} Galaxy distribution in the most and least massive 25
        clusters (the mean masses are shown in the figure).
        The galaxy radial distributions are consistent
        with each other. The two profiles are fit with
        $c_g = 2.8\pm0.3$ (high mass) \& $2.9\pm0.5$ (low mass).
        We adjust the relative normalization of
        the profiles for ease of comparison.
     }
\end{inlinefigure}

In Fig~\ref{fig:cfprof} (lower panel) we show the profiles for stacked clusters
built from the 25 most and least massive clusters (again excluding their BCGs). 
The normalization of the profiles is arbitrarily chosen so the two profiles are 
clearly displayed.  These two radial distributions are statistically consistent,
indicating that groups and clusters have similar radial galaxy distributions.
Fitting the NFW model to the two, the 
best-fit concentration parameters are $2.84_{-0.32}^{+0.37}$ \& 
$2.88_{-0.52}^{+0.55}$, for high mass and low mass clusters, respectively. 
Therefore, any variation of galaxy concentration with respect to cluster mass
must be weak.

Our finding indicates that, 
whatever the mechanisms that are responsible for the galaxy population 
transformation, they have to operate in such a way that the galaxy radial 
distribution is unaffected. 
This suggests that mechanisms whose efficiency strongly depends on
the typical clustercentric distance of a galaxy are not preferred. Thus, it may 
be that ram pressure stripping and mean field tidal stripping are not major 
contributors to the population difference in high and low mass clusters, unless 
the bulk of galaxies are on highly eccentric orbits that take them near the 
cluster center.

By combining several high-resolution simulations, \citet{delucia04}
are able to address the subhalo radial distribution in parent haloes
of different masses, under the effects of tidal stripping.  They find
that the (3D) subhalo number density is similar in host haloes of mass
$10^{14}$ and $10^{15} M_\odot$ (their figure 6), which is in good agreement 
with our finding.  However, they also find that more massive subhaloes tend to
reside at larger radii from the center of host haloes, because of the
efficiency of tidal stripping within the cluster environment. In 
Fig~\ref{fig:cfprof} (upper panel) we 
examine the distribution of bright and faint galaxies (with an
arbitrary luminosity separation at $M_K=-23.5$) and find that the two
galaxy populations have similar projected radial profiles (the bright galaxy
distribution has $c_g = 2.99_{-0.25}^{+0.31}$, while the faint galaxy 
distribution has
$c_g = 2.75^{+0.33}_{-0.32}$). We note that when the BCGs are included,
the profile of bright galaxies becomes very cuspy and thus quite different from 
that of faint galaxies; however, \citet{delucia04}
do not include the BCGs in their sample either 
(which are identified as the friend-of-friend groups themselves).
If the mass of a subhalo has a simple correlation with the luminosity of
the galaxy that resides inside, the simulation result appears to be inconsistent
with our analysis.
A further disagreement appears when comparing the galaxy distribution directly
with that
of subhaloes in numerical simulations \citep{diemand04}; subhaloes have a much 
less concentrated distribution.

The results in these two sections seem to pose some difficulties for dynamical
processes. Tidal interactions surely can cause mass loss in subhaloes, and 
may produce the observed galaxy profile (assuming there is an one-to-one 
correspondence between a subhalo and a galaxy). However, the majority of 
galaxies that need to be absent in high mass clusters (those fainter than $M_*$)
may not be easily destroyed by tides. In addition, the similar spatial distribution of
galaxies (in high and low mass clusters and for bright and faint galaxies)  provides some argument
against tidal interaction, ram pressure stripping, and cannibalism by the central galaxy.

\subsection{Redshift Evolution of the HOD}

Comparison of present epoch groups and clusters may lead to a biased picture of the transformation of the galaxy population.
The progenitors of the present day high mass clusters, namely the lower
mass clusters at higher redshifts, may not have the same properties as
present day low mass systems. It may be that the moderate or low mass
clusters at high redshift have systematically fewer galaxies per unit mass than their present epoch counterparts. 
Interestingly, high resolution $N$-body numerical studies have shown just the opposite; the
normalization of the halo occupation number of subhaloes increases by
a factor of 3 from $z=0$ to $z=5$  \citep{kravtsov03}.  

A clean way to examine this would be to trace the cluster LFs to high
redshifts and compare clusters of roughly the same mass. An existing
dataset makes this comparison possible \citep[][a heterogeneous sample
of 38 clusters from $0.1 < z <1$]{depropris99}; each cluster in the
sample is imaged in $K_s$-band to about $M_*(z)+2$ mag (assuming passive
evolution in galaxy luminosity). We collect the emission-weighted mean
temperature $T_X$ from literature for 20 of these clusters, and we 
re-scale the
background-corrected number of galaxies to $r_{500}$ for the clusters
with $T_X$ information.  $r_{500}$ is estimated by assuming the
standard evolution of the $M$--$T_X$ relation
\citep[e.g.][]{mohr00a}. Using the best-fit values for $M_*$ at
different cosmic epochs, we estimate the number of galaxies to a depth
two magnitudes below $M_*$ for each cluster in the high-$z$ sample.
We also calculate the halo occupation number for our local sample in
the same way. By doing so we hope to reduce any systematic differences
between the low and high redshift cluster samples.

\begin{inlinefigure}
   \ifthenelse{\equal{\figtype}{EPS}}{
   \begin{center}
   \epsfxsize=8.cm
   \begin{minipage}{\epsfxsize}\epsffile{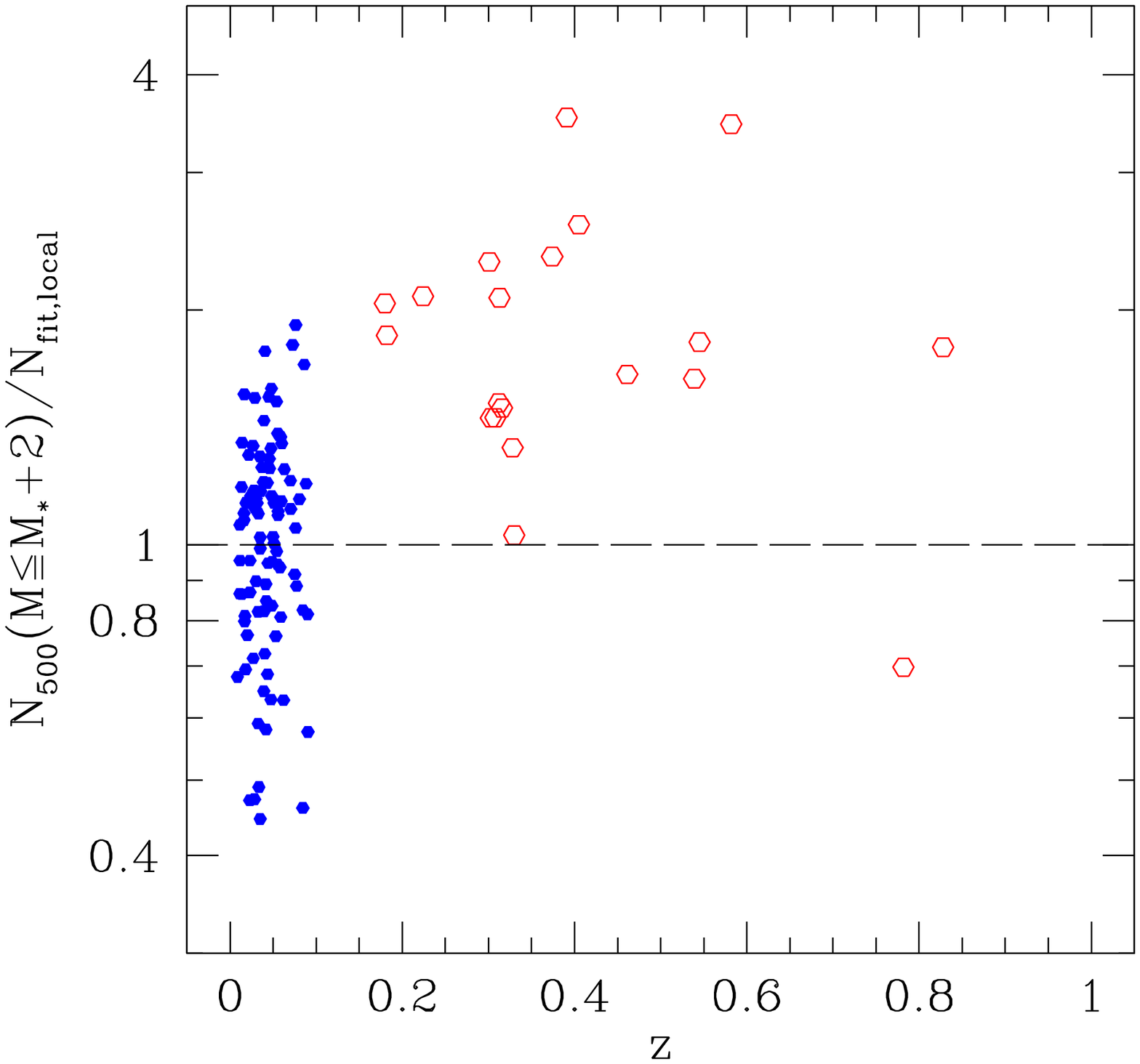}\end{minipage}
   \end{center}}
   {\myputfigure{f9.pdf}{0.0}{1.0}{-70}{-40}}
   \figcaption{\label{fig:depropris}
        Comparison of halo occupation number for our local sample (solid points)
        and the distant clusters from the sample of
        \citet[][hollow points]{depropris99}.
        We plot the ratio between the derived galaxy number and the best-fit
        $N(M)$ for each cluster in the two samples. The best-fit relation is
        obtained from the local sample with cutoff magnitude at $M_K=-22$,
        roughly two magnitudes fainter than $M_{K*}$ (c.f. Eqn~\ref{eq:hod},
        \S\ref{sec:clf}).
     }
\end{inlinefigure}

This exercise shows that the clusters at high redshift have a larger
halo occupation number, compared to the clusters of similar mass in
our low-$z$ sample. The results are shown in Fig~\ref{fig:depropris}, where we
plot the ratio of the total galaxy number to the expected number in a local cluster of the same mass
(using the best-fit $N$--$M$ relation of the nearby sample).
As expected, the nearby clusters scatter about unity. However,
the clusters at higher redshifts almost all have a larger number of 
galaxies. The result holds true when the comparison is made at a smaller
radius $r_{1800}$, where less extrapolation for the high-$z$ sample is needed.
Although we attempt to match the analyses of the local 2MASS and high-$z$ samples, 
systematic effects such as differences in star-galaxy separation in the two 
catalogs may still contribute to the differences in properties. 
However, if the high-$z$ clusters do
have a larger halo occupation number, our comparison then supports the
finding of \citet{kravtsov03}, and rules out the possibility that progenitors of
present day high-mass clusters have smaller mean galaxy number density compared
to that of the low mass clusters we see today.  In other words, the redshift 
evolution of the HOD appears to go in the wrong direction to explain the 
flatness of the local $N$--$M$ relation. Observations have suggested a higher
frequency of merger events in high redshift clusters \citep[e.g.][]{vandokkum00}. 
It is therefore likely that the higher mean occupation number found in high-$z$
clusters is reduced over cosmic time as mergers take place.

To improve our understanding of the evolution of the HOD,  we have begun an observational program to improve the available data on the local and high-$z$ samples.  In addition, a {\it Spitzer Space Telescope} GTO program will sample the rest-frame $K$-band light from distant clusters, 
which should provide more direct constraints on this aspect of cluster 
evolution.

\section{Systematic Uncertainties}
\label{sec:system}

Before we consider the implications of our findings on the hierarchical
structure formation, it is important to examine the robustness of our results. 
Here we investigate several factors that may potentially limit the 
interpretation of our results. In the Appendix we further consider the effects
of our cluster selection and background subtraction method on the scaling 
relations.

{\it Galactic Contamination}. To obtain a large sample, we choose 
objects 10 deg out of the Galactic plane; however, the possible confusion from
stars may bias our results for clusters located at $|b| \le 25^\circ$ 
\citep[e.g.][]{bell03}.
We therefore re-examine the scaling relations for the 78 clusters far from the
Galactic plane ($|b|>25^\circ$). The results remain unchanged.

{\it Uncertainties in Cluster Radius and Center.} Fitting the galaxy
distribution requires knowledge of the virial
radius and the cluster center for each cluster. To test the
robustness of our fit of the galaxy profile, we build mock clusters with 
characteristics similar to those in our sample (mass, redshift, halo occupation
number, etc) and stack them,
with uncertainties in the virial radius (up to $20\%$) and cluster center (up to $5'$)
added.  Without any uncertainties, our fitting routine gives a slightly positive
biased result ($\Delta c \equiv c_{fit}-c_{input} \sim 0.06$ with scatter about 0.24).
Uncertainties in the virial radius tend to cause a larger concentration than
the true value ($c_{fit}>c_{input}$), while uncertainties in cluster center
act in the opposite direction. For example, a $10\%$ uncertainty in the virial
radius gives $\Delta c \sim 0.26$ with a scatter of 0.25, and a $1'$
uncertainty in cluster center results in $\Delta c \sim -0.06$ and a
similar scatter. For probable uncertainties in the above two quantities ($10-20\%$ \&
$\sim 1'$), we find that $\Delta c \sim 0.15 - 0.35$, with scatter about 0.25.
Therefore, it is 
possible that the true galaxy distribution concentration is even smaller than 
2.9 found in \S\ref{sec:profile}. However, considering the fact that the LFs 
evaluated at $r_{500}$ \& 
$r_{200}$ also suggest $c_g \sim 3$, we believe our result should not be too
different from the true value. 

{\it Shape of the Luminosity Function}. The choice of $\alpha$ is
important, especially in deriving the total number of galaxies, and
thus the halo occupation number. As we have shown in \S\ref{sec:clf}, the 
value we have adopted is consistent with the stacked luminosity function from our cluster sample.
We test the scaling relations with different choices of the faint-end slope:
assuming $\alpha=-0.9$ or $-1.3$ gives very similar (within $1\sigma$) slopes
for light-mass correlation, while resulting slopes of halo occupation number
that are $\sim 1.5\sigma$ away (and biased low and high, respectively) from the 
values presented for $\alpha=-1.1$.

{\it Mass--Temperature Relation}. The $M$--$T_X$ relation (Eqn~\ref{eq:mt})
we adopt for 
estimating cluster mass from \xray temperature is mostly suitable for systems
more massive than $M_{500} = 3.57\times 10^{13} h_{70}^{-1} M_\odot$
\citep{finoguenov01}. Although nearly all our clusters are more massive than 
this mass scale, it is important to
consider the effect of a different $M$--$T_X$ relation on the scaling relations
found in this paper. Using a steeper $M$--$T_X$ relation (e.g. the one with
slope of 1.78 but a smaller normalization, see Table 1 of 
\citealt{finoguenov01}) 
would result in slightly flatter $L$--$M$ \& $N$--$M$ relations whose slopes are
about $1-2 \sigma$ smaller than our best-fit values. 

{\it Cluster Mass Dependence of Matter and Galaxy Distribution}. Numerical
simulations suggest that the matter distribution concentration $c_{dm}$ has a
weak dependence on halo mass \citep{bullock01}. Because galaxies may generally 
follow the dark matter, it is also possible that $c_g$ is a function of cluster 
mass. Because $r_{500}$ is inferred directly from \xray observations, while 
$r_{200}$ is converted by assuming $c_{dm}=5$, the relations evaluated at 
$r_{200}$ may be affected by varying $c_{dm}(M)$. Adopting the prescription for 
$c_{dm}(M)$ given by \citet{klypin99} yields an almost identical 
$L_{200}$--$M_{200}$ relation.  Systematic variation of $c_g(M)$ would come into
our analysis through the conversion from the observed, ``projected'' quantities
to those contained within the cluster spherical volume (see 
\S\ref{sec:profile}).  As we show in \S\ref{sec:profconstraint}, 
the data provide no indication of mass related trends in $c_g$.

These tests indicate that our results are not sensitive to the parameters or
methods that we adopt.

\section{Conclusions}
\label{sec:summary}

We have analyzed a sample of 93 galaxy clusters whose virial masses range
from $3\times 10^{13} h_{70}^{-1} M_\odot$ to $1.7\times 10^{15} h_{70}^{-1} 
M_\odot$.   We have used the peak in X-ray surface brightness to define the 
cluster centers, and we have used the observed correlation between 
emission-weighted mean temperature and cluster virial mass \citep{finoguenov01} 
along with published cluster redshifts to estimate the scale of the cluster 
virial region.  The following
bulk properties of these systems are studied, with emphasis on any mass related
trends: the composite cluster luminosity function, the galaxy distribution 
within the clusters, the correlation between total galaxy $K$-band luminosity 
and the cluster binding mass, and the halo occupation number.

Our analysis shows that the luminosity function (excluding the BCG) is well 
fit by a Schechter function with a faint-end slope between $-1.1 \lesssim 
\alpha \lesssim -0.84$, and characteristic magnitude and number 
density between $-24.34 \le M_* -5\log h_{70} \le -24.02$ and $3.01\,h_{70}^3$ Mpc$^{-3} \le \phi_* 
\le 4.43\,h_{70}^3$ Mpc$^{-3}$.   A NIR survey of nearby clusters to a
depth of $M_K \sim -19$ would be useful in further constraining the faint-end slope.
The composite galaxy distribution ($0.02 \le r/r_{200} \le 2$) 
is found to be an NFW profile with concentration
$c_g = 2.9\pm 0.2$. This is somewhat smaller than 
the observed dark matter  concentration, as well as that found in numerical 
studies. This implies that galaxies have a more extended distribution than that 
of the dark matter.

By solving the LF parameters $L_*$ and $\phi_*$ for each cluster, we obtain the
total galaxy luminosity, as well as total galaxy number (\S\ref{sec:lmhod}) above some limiting absolute magnitude $M_{K,low}=-21$.
By comparing the galaxy NIR light--cluster mass correlation at different radii ($r_{500}$ \&
$r_{200}$) we find that the \mlr decreases slightly with
radius (as would be expected given the differing distributions of galaxy and dark matter).  
The $L$--$M$ correlation has a slope smaller than unity 
($0.69\pm 0.04$, see Eqn \ref{eq:lm}, Table 1), which 
implies that cluster \mlr is an increasing function of cluster mass. However,
excluding the BCGs significantly affects the normalization and the slope of the 
$L$--$M$ correlation (but of course has only a minor effect on the halo 
occupation number). Without the BCGs, the total galaxy light roughly scales as 
the total galaxy number, which increases more rapidly with cluster mass 
(Eqn~\ref{eq:hod}, slope is $0.84\pm0.04$; Table 1). The slope of the $N$--$M$ 
correlation agrees well with the prediction of numerical simulations. In
addition, we find the distribution of the halo occupation number about the
mean is Poissonian (Figure~\ref{fig:hod}).

This flatness of the $N$--$M$ relation in our local cluster sample indicates 
that higher mass clusters have fewer galaxies per unit mass than low mass 
clusters.    
We examine the occupation number of higher redshift clusters using published NIR and X-ray data; in crude agreement with numerical simulations, the higher redshift clusters have more galaxies per unit mass than nearby clusters.  
As massive clusters accrete lower mass systems, there must be  processes at work that alter the galaxy populations, either reducing galaxy masses or perhaps even destroying some galaxies.  
Through further examination of our local cluster sample we find that the radial distribution of galaxies appears to be independent of cluster mass; in addition the ultra bright end of the luminosity function is simply not present within low mass clusters and groups, and the number density of galaxies fainter than about $M_*$ is higher in low mass clusters than in high mass clusters.  
Thus, the transformation of the galaxy population from low mass to high mass clusters appears to be a building up of a few extremely luminous galaxies and the destruction or faintening of galaxies fainter than $M_*$.  
We note that there does not appear to be enough light in missing galaxies in 
high mass clusters to 
account for the increases in the bright end of the luminosity function.

We investigate possible mechanisms that shape the slope of the halo
occupation $N$--$M$ correlation (\S\ref{sec:further}).  We find that none of the mechanisms
we consider can by itself offer a satisfactory explanation of the galaxy population 
differences between high and low mass clusters.  Among these, tidal interactions
(stripping, galaxy harassment) could be quite important,  but presumably merging must also play a role in the production of very luminous galaxies.
Further theoretical progress together with additional observational constraints from extensions of studies like ours will presumably lead to a clearer picture of the process of galaxy evolution within the cluster environment.

One implication of the observed $N$--$M$ relation in both our local sample and the high-$z$ sample is that galaxies are destroyed or dramatically altered through various dynamical processes in the cluster environment. One thus expects that the amount of the diffuse intracluster light would exhibit some correlation with the cluster mass \citep[e.g.][]{malumuth84}. 
In a companion paper (paper III), we will study the 
relationship between the BCG luminosity and the cluster mass, using it to limit the amount of the intracluster diffuse light. This will provide further insights into the cluster formation scenarios and the evolution of cluster galaxies.

An improved understanding of  the redshift evolution of the
HOD is critical for those who would use deep, NIR or optical cluster surveys
to study cosmology.  High yield cluster surveys must employ simple mass estimators such as the galaxy number or luminosity, because it would not be feasible to conduct detailed mass measurements using X-ray imaging spectroscopy, galaxy velocity dispersions or weak lensing.
Our study provides a local calibration for these mass--observable 
relations. However, extracting cosmology from the observed cluster redshift distribution will require knowledge of how these relations evolve.   
Our comparison of the local sample to the published data on the high-$z$ clusters suggests that cluster virial regions are better represented by galaxy number and light at high redshift than they are locally.  
Depending on how the non-cluster galaxies are evolving, this could boost optical
and NIR cluster finding at high redshift.  It remains to be seen whether the 
recently discussed cluster survey self-calibration approach 
\citep{majumdar03b,hu03b} will work with the higher scatter galaxy-based mass 
indicators.

\acknowledgements

We thank an anonymous referee for comments that improve the paper.
We acknowledge Andrey Kravtsov and Stefano Andreon for many helpful suggestions.
We thank Tom Jarrett for providing the 2MASS all-sky $\log N - \log S$ relation
prior to publication.  YTL thanks Al Sanderson and I. H. for helpful 
discussions.
SAS acknowledges support from the NASA Long Term Space Astrophysics grant NAG 
5-8430.  The work by SAS at Lawrence Livermore National Laboratory was 
performed under the auspices of the Department of Energy under Contract 
W-7405-ENG-48.
This publication makes use of data products from the Two Micron All Sky Survey, 
which is a joint project of the University of Massachusetts and the Infrared 
Processing and Analysis Center/Caltech, funded by 
the NASA and the NSF.
This research has made use of the NASA/IPAC Extragalactic Database 
(NED), 
and the X-Ray Clusters Database (BAX).

\appendix

Here we provide some tests that help assess the robustness of the
scaling relation between the cluster light and mass presented in this paper.
We show below that the use of the statistical background subtraction method is 
adequate and does not cause noticeable mass-related trends. Further we show the
use of a heterogeneous cluster sample does not bias the $L$--$M$ and $N$--$M$
scaling relations.

{\it Statistical Background Subtraction.} We first probe for mass related biases
in the statistical background subtraction by examining the fraction of the
observed flux that comes from the background. This may be a concern
since for more massive clusters (which tend to be more distant) the total 
light derived may be more sensitive to the reliability of the statistical 
background method. Fig A1 (left panel)
shows there is no obvious trend related to cluster mass. Typically 16\% of the 
observed flux is subtracted as background flux. For the
78 clusters that do not require large statistical background flux 
subtraction ($f_{bgn}/f_{obs}\le 0.23$), we find the slope of the $L$--$M$
relation remains the same.

Furthermore, a comparison of the background galaxy counts estimated by the
statistical background method and by the ``annulus'' background method can 
reveal how secure the former method is against field-to-field variations.
Under the assumption that cluster galaxies are distributed according to an
NFW profile with $c_g=3$, we solve for the constant background $\Sigma_{b,a}$
and the overall normalization of the NFW profile by the total galaxy counts
in cluster virial region and within a large annulus around the cluster (c.f.
\S 2.2.1 in paper I). We show the resultant ``annulus'' background surface 
density in Fig A1 (right panel). 
Given the large amplitude of the angular correlation function of the 2MASS
galaxies \citep{maller03}, which in turn implies a non-negligible
contribution of background galaxy clustering (similar or larger than the
Poisson error, e.g. \citealt{peebles80,blake02}), the large field-to-field 
variation is not surprising.  However, we note that
the weighted (unweighted) average $\overline{\Sigma_{b,a}} = 12.0\pm0.5$
($15.3\pm1.1$) deg$^{-2}$ is in reasonable agreement with the expectation of
$\Sigma_{b,s} = 14.4 \pm 0.3$ deg$^{-2}$, obtained by
integrating the $\log N$--$\log S$ to the 2MASS completeness limit $K=13.5$
(shown in the figure as a horizontal dashed line).
Note that the uncertainty in $\Sigma_{b,s}$ only accounts for the uncertainty in
the $\log N$--$\log S$ relation.  It is interesting 
that for two clusters, $\Sigma_{b,a} <0$, which probably indicates that an NFW
profile does not describe the galaxy distribution well. Finally, we
note that the clusters are sorted in an order of decreasing mass; it thus 
appears that employing the statistical background method does not introduce any 
mass trends.

\begin{figure}
\figurenum{A1}
\epsscale{0.8}
\plottwo{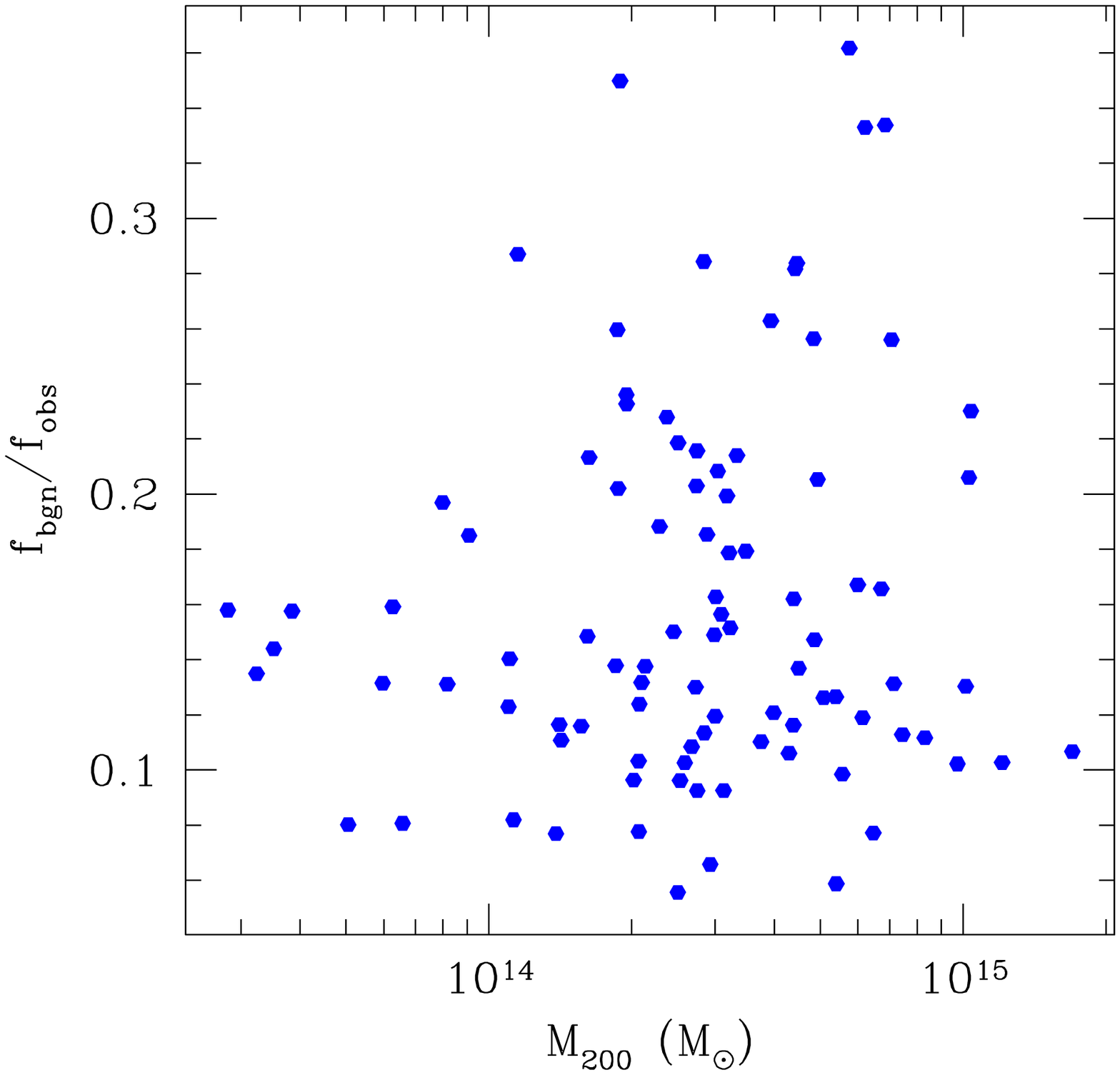}{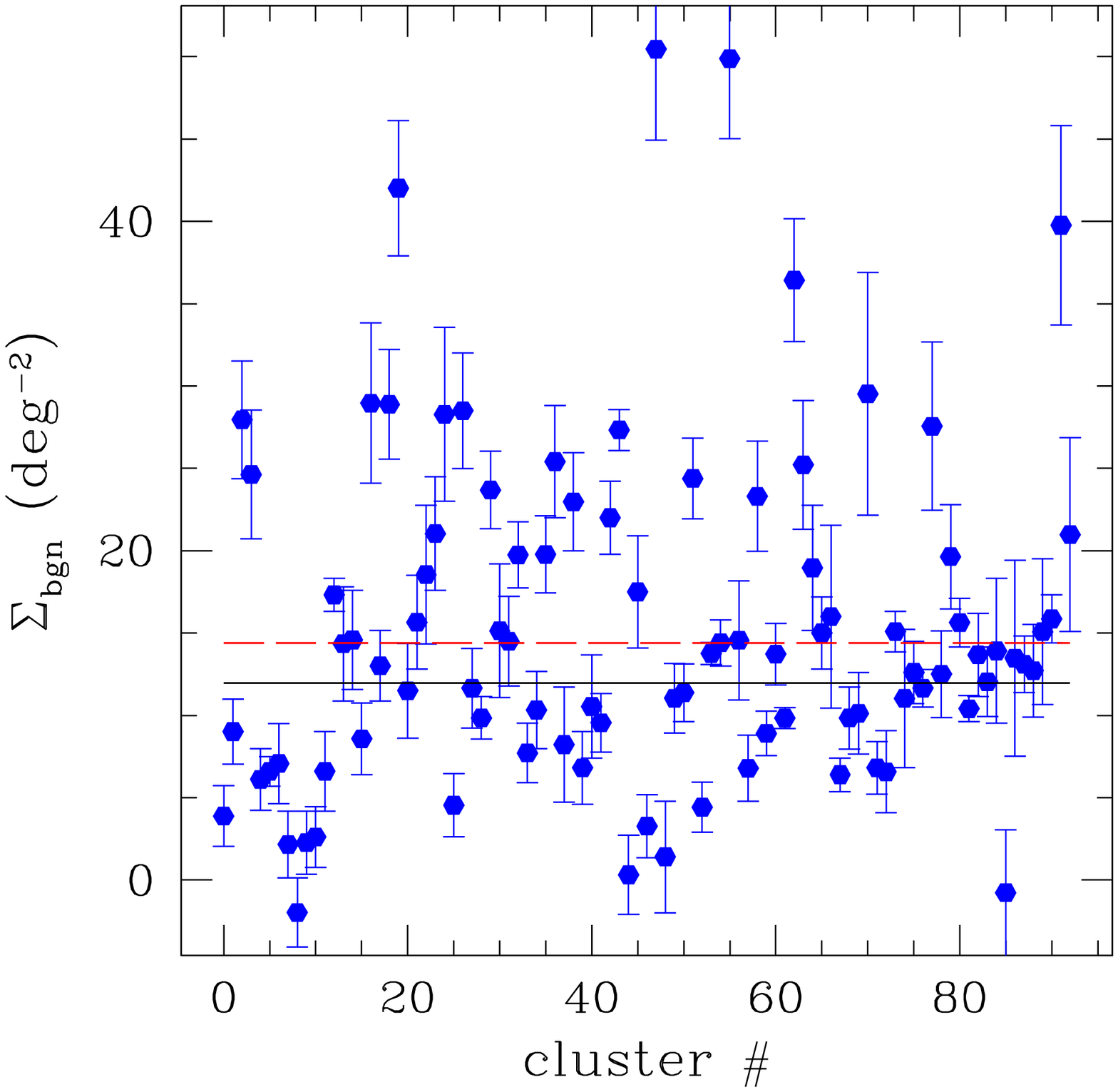}
\caption{
    {\it Left}: ratio of the estimated background flux to the total
    observed flux, as a function of cluster mass. {\it Right}: comparison
    between the ``annulus'' background surface densities (points) and that
    estimated from the all-sky $\log N$--$\log S$ relation (dashed line,
    $\Sigma_{b,s} = 14.4 \pm 0.3$ deg$^{-2}$).
    The solid line is the weighted average value of the points 
    ($\overline{\Sigma_{b,a}} = 12.0 \pm 0.5$ deg$^{-2}$).}
\end{figure}

{\it Cluster Selection}. 
As we point out previously, our cluster sample is basically X--ray-selected;
more massive clusters tend to be at higher redshifts, and therefore any 
redshift-related systematics may be interpreted as a mass trend.
Following paper I, we divide our clusters into ``hot'' and ``cold'' subsamples, 
with $kT_X=3.8$ keV as the dividing point. The mean redshifts
for the hot and cold subsamples are 0.056 and 0.031, respectively.
To investigate if any redshift trend exists, we choose $z=0.043$ 
to separates clusters into ``nearby'' and ``distant'' categories. In Fig A2 
we show the LF parameters $M_*$ and $\phi_*$. The filled pentagons
represent hot clusters, while open circles denote cold clusters. Furthermore,
the relative sizes of the symbols indicate whether they are nearby or distant.
Dividing the whole sample into distant and nearby subsamples, a two-dimensional
Kolmogorov-Smirnov test indicates that the two subsamples are highly likely to 
be drawn from the same parent distribution. Within each 
temperature group, there is also no difference between $M_*$ and $\phi_*$ for 
distant and nearby clusters.

\begin{figure}
\figurenum{A2}
\epsscale{0.38}
\plotone{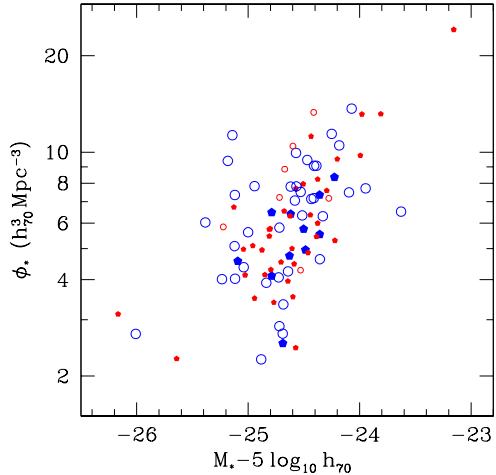}
\caption{
    The LF parameters for individual clusters. Filled pentagons:
    clusters with $kT_X>3.8$ keV, open circles: those with $kT_X \le 3.8$ keV.
    Larger symbols: clusters at $z\le 0.043$, smaller symbols: those at $z>
    0.043$.
    }
\end{figure}

Because 2MASS is a flux-limited survey, only the brightest galaxies can be 
observed in more distant clusters. Therefore deriving the LF
for these distant-- approaching $z\sim0.1$-- clusters (especially when they are 
not massive) requires some extrapolation. We have examined the scaling 
relations by excluding those 
clusters whose estimated total light is more than 3 times the total observed 
light. The changes in the slopes of the scaling relations are within the 
statistical uncertainties. Finally, we consider a subsample of 49 clusters
that are within HIFLUGCS \citep{reiprich02}, which is a \xray flux-limited 
sample. The scaling relations based on
this subsample show slight increases in slope, but all within $1\sigma$ from 
those listed in Table 1.


\begin{table*}[htb]
\begin{center}
\caption{Derived Quantities of the Clusters}
{\scriptsize
\begin{tabular}{lclllcclcl}
\tableline \tableline
      &      & $T_X$ &            &  $\phi_{*,500}$              & $M_{*}-5\log h_{70}$ &  $L_{500}$                      &  & $L_{200}$ & \\
Name  & $z$  & (keV) &  $N_{obs}$ &($h_{70}^3\,$Mpc$^{-3}$)& (mag) &($h_{70}^{-2}\,10^{12} L_\odot$) & $N_{500}$ & ($h_{70}^{-2}\,10^{12} L_\odot$) & $N_{200}$ \\
  (1) & (2)  & (3)   &  (4)       & (5)                    & (6)   & (7)                             & (8)  & (9) & (10)  \\
\tableline
a2319 & 0.0557 & $11.8^{0.09}_{0.09}$ & 74 & $4.54 \pm 0.41$ & $-24.70 \pm 0.23$ & $14.63 \pm 0.04$ & $267 \pm 32$ & 	$22.17 \pm 0.03$ & $399 \pm 46$ \\ 
trian & 0.0510 & $9.5^{0.42}_{0.42}$ & 62 & $4.14 \pm 0.43$ & $-24.85 \pm 0.26$ & $12.01 \pm 0.05$ & $184 \pm 25$ & 	$18.01 \pm 0.04$ & $283 \pm 35$ \\ 
a2029 & 0.0773 & $8.7^{0.18}_{0.18}$ & 14 & $5.30 \pm 1.38$ & $-24.22 \pm 0.55$ & $8.12 \pm 0.06$ & $155 \pm 44$ & 	$13.14 \pm 0.05$ & $277 \pm 76$ \\ 
a2142 & 0.0899 & $8.68^{0.12}_{0.12}$ & 12 & $3.96 \pm 1.01$ & $-24.64 \pm 0.58$ & $7.20 \pm 0.09$ & $134 \pm 38$ & 	$13.38 \pm 0.06$ & $265 \pm 70$ \\ 
a0754 & 0.0542 & $8.5^{0.3}_{0.3}$ & 47 & $4.30 \pm 0.50$ & $-24.80 \pm 0.30$ & $9.28 \pm 0.05$ & $156 \pm 24$ & 	$12.79 \pm 0.04$ & $232 \pm 35$ \\ 
a1656 & 0.0232 & $8.21^{0.16}_{0.16}$ & 122 & $4.10 \pm 0.38$ & $-24.79 \pm 0.26$ & $8.94 \pm 0.05$ & $147 \pm 18$ & 	$12.58 \pm 0.05$ & $220 \pm 25$ \\ 
a2256 & 0.0601 & $7.51^{0.11}_{0.11}$ & 45 & $6.55 \pm 0.69$ & $-24.68 \pm 0.29$ & $10.11 \pm 0.04$ & $186 \pm 26$ & 	$14.27 \pm 0.04$ & $265 \pm 38$ \\ 
a3667 & 0.0560 & $7^{0.36}_{0.36}$ & 41 & $4.95 \pm 0.60$ & $-24.88 \pm 0.31$ & $8.65 \pm 0.06$ & $136 \pm 22$ & 	$12.76 \pm 0.04$ & $233 \pm 34$ \\ 
a2255 & 0.0806 & $6.87^{1.7}_{1.1}$ & 21 & $5.10 \pm 0.74$ & $-24.96 \pm 0.41$ & $8.70 \pm 0.08$ & $135 \pm 26$ & 	$12.28 \pm 0.08$ & $185 \pm 36$ \\ 
a0478 & 0.0881 & $6.84^{0.13}_{0.13}$ & 11 & $6.32 \pm 1.36$ & $-24.62 \pm 0.58$ & $7.96 \pm 0.09$ & $146 \pm 35$ & 	$12.95 \pm 0.06$ & $321 \pm 117$ \\ 
a1650 & 0.0845 & $6.7^{0.5}_{0.5}$ & 5 & $2.45 \pm 1.21$ & $-24.57 \pm 1.01$ & $3.09 \pm 0.14$ & $54 \pm 29$ & 	$6.97 \pm 0.07$ & $172 \pm 86$ \\ 
a0644 & 0.0704 & $6.59^{0.1}_{0.1}$ & 19 & $6.37 \pm 1.08$ & $-24.44 \pm 0.45$ & $6.34 \pm 0.06$ & $133 \pm 28$ & 	$9.05 \pm 0.06$ & $166 \pm 37$ \\ 
a0426 & 0.0183 & $6.33^{0.21}_{0.18}$ & 120 & $4.57 \pm 0.44$ & $-25.09 \pm 0.29$ & $8.63 \pm 0.07$ & $122 \pm 15$ & 	$12.12 \pm 0.05$ & $205 \pm 21$ \\ 
a1651 & 0.0860 & $6.3^{0.3}_{0.3}$ & 10 & $13.14 \pm 4.28$ & $-23.98 \pm 0.64$ & $7.82 \pm 0.05$ & $205 \pm 63$ & 	$12.05 \pm 0.04$ & $383 \pm 197$ \\ 
a3266 & 0.0594 & $6.2^{0.5}_{0.4}$ & 32 & $5.47 \pm 0.72$ & $-24.81 \pm 0.35$ & $7.79 \pm 0.06$ & $121 \pm 21$ & 	$11.71 \pm 0.04$ & $217 \pm 34$ \\ 
a0085 & 0.0551 & $6.1^{0.12}_{0.12}$ & 27 & $5.00 \pm 0.76$ & $-24.61 \pm 0.39$ & $5.99 \pm 0.06$ & $101 \pm 20$ & 	$8.02 \pm 0.05$ & $156 \pm 32$ \\ 
a2420 & 0.0846 & $6^{2.3}_{1.2}$ & 8 & $4.86 \pm 1.61$ & $-24.46 \pm 0.75$ & $4.71 \pm 0.10$ & $86 \pm 32$ & 	$8.40 \pm 0.06$ & $241 \pm 140$ \\ 
a0119 & 0.0440 & $5.8^{0.36}_{0.36}$ & 40 & $4.13 \pm 0.56$ & $-25.03 \pm 0.34$ & $6.46 \pm 0.07$ & $91 \pm 16$ & 	$10.10 \pm 0.08$ & $133 \pm 21$ \\ 
a3391 & 0.0531 & $5.7^{0.42}_{0.42}$ & 23 & $3.50 \pm 0.61$ & $-24.94 \pm 0.44$ & $5.84 \pm 0.09$ & $72 \pm 16$ & 	$9.29 \pm 0.07$ & $132 \pm 25$ \\ 
a3558 & 0.0480 & $5.7^{0.12}_{0.12}$ & 60 & $6.73 \pm 0.68$ & $-25.13 \pm 0.26$ & $11.10 \pm 0.06$ & $147 \pm 20$ & 	$17.98 \pm 0.05$ & $260 \pm 28$ \\ 
a3158 & 0.0590 & $5.5^{0.3}_{0.4}$ & 31 & $7.70 \pm 0.98$ & $-24.57 \pm 0.35$ & $6.76 \pm 0.05$ & $129 \pm 22$ & 	$9.13 \pm 0.05$ & $169 \pm 30$ \\ 
a1991 & 0.0590 & $5.4^{5.9}_{2.2}$ & 17 & $4.48 \pm 0.90$ & $-24.59 \pm 0.50$ & $4.06 \pm 0.07$ & $73 \pm 19$ & 	$5.47 \pm 0.06$ & $108 \pm 32$ \\ 
a2065 & 0.0726 & $5.37^{0.206}_{0.181}$ & 23 & $11.20 \pm 1.41$ & $-24.44 \pm 0.38$ & $7.33 \pm 0.05$ & $167 \pm 28$ & 	$12.40 \pm 0.03$ & $329 \pm 60$ \\ 
a1795 & 0.0631 & $5.34^{0.07}_{0.07}$ & 17 & $9.76 \pm 1.90$ & $-23.99 \pm 0.48$ & $4.80 \pm 0.04$ & $123 \pm 29$ & 	$6.25 \pm 0.04$ & $182 \pm 58$ \\ 
a3822 & 0.0760 & $5.12^{0.26}_{0.187}$ & 16 & $4.97 \pm 0.86$ & $-25.04 \pm 0.48$ & $5.38 \pm 0.10$ & $85 \pm 19$ & 	$9.36 \pm 0.07$ & $165 \pm 34$ \\ 
a2734 & 0.0620 & $5.07^{0.36}_{0.42}$ & 11 & $3.53 \pm 0.94$ & $-24.60 \pm 0.64$ & $3.13 \pm 0.09$ & $53 \pm 17$ & 	$4.11 \pm 0.07$ & $89 \pm 36$ \\ 
a3395sw & 0.0510 & $5^{0.3}_{0.3}$ & 30 & $5.77 \pm 0.81$ & $-24.80 \pm 0.37$ & $5.61 \pm 0.06$ & $92 \pm 17$ & 	$7.98 \pm 0.06$ & $134 \pm 24$ \\ 
a0376 & 0.0484 & $5^{2}_{1.1}$ & 29 & $7.59 \pm 1.10$ & $-24.30 \pm 0.37$ & $4.57 \pm 0.04$ & $100 \pm 19$ & 	$6.49 \pm 0.03$ & $171 \pm 33$ \\ 
a1314 & 0.0335 & $5^{4.5}_{1.8}$ & 22 & $2.53 \pm 0.64$ & $-24.69 \pm 0.58$ & $2.48 \pm 0.09$ & $40 \pm 13$ & 	$2.60 \pm 0.16$ & $38 \pm 20$ \\ 
a2147 & 0.0351 & $4.91^{0.18}_{0.18}$ & 42 & $5.76 \pm 0.80$ & $-24.50 \pm 0.35$ & $4.15 \pm 0.05$ & $81 \pm 15$ & 	$7.58 \pm 0.04$ & $150 \pm 22$ \\ 
a3112 & 0.0750 & $4.7^{0.24}_{0.24}$ & 8 & $6.00 \pm 1.73$ & $-24.38 \pm 0.71$ & $4.11 \pm 0.08$ & $71 \pm 24$ & 	$6.44 \pm 0.06$ & $150 \pm 57$ \\ 
a1644 & 0.0474 & $4.7^{0.9}_{0.7}$ & 34 & $7.96 \pm 1.03$ & $-24.51 \pm 0.34$ & $5.82 \pm 0.05$ & $103 \pm 18$ & 	$7.91 \pm 0.04$ & $150 \pm 25$ \\ 
a2199 & 0.0303 & $4.5^{0.2}_{0.1}$ & 52 & $7.35 \pm 0.93$ & $-24.36 \pm 0.32$ & $4.43 \pm 0.04$ & $86 \pm 15$ & 	$6.88 \pm 0.04$ & $138 \pm 20$ \\ 
a2107 & 0.0421 & $4.31^{0.57}_{0.35}$ & 23 & $5.54 \pm 1.03$ & $-24.36 \pm 0.46$ & $3.26 \pm 0.05$ & $60 \pm 15$ & 	$4.38 \pm 0.07$ & $76 \pm 19$ \\ 
a0193 & 0.0486 & $4.2^{1}_{0.5}$ & 17 & $5.45 \pm 1.13$ & $-24.39 \pm 0.52$ & $3.16 \pm 0.06$ & $57 \pm 15$ & 	$4.05 \pm 0.05$ & $83 \pm 24$ \\ 
a2063 & 0.0355 & $4.1^{0.6}_{0.6}$ & 38 & $8.36 \pm 1.16$ & $-24.23 \pm 0.35$ & $3.50 \pm 0.04$ & $79 \pm 15$ & 	$4.53 \pm 0.04$ & $107 \pm 21$ \\ 
a4059 & 0.0475 & $4.1^{0.18}_{0.18}$ & 14 & $3.40 \pm 0.83$ & $-24.77 \pm 0.60$ & $3.12 \pm 0.10$ & $39 \pm 12$ & 	$4.71 \pm 0.07$ & $78 \pm 20$ \\ 
a1767 & 0.0701 & $4.1^{1.8}_{0.9}$ & 11 & $8.23 \pm 1.77$ & $-24.38 \pm 0.58$ & $4.20 \pm 0.07$ & $79 \pm 21$ & 	$5.36 \pm 0.07$ & $99 \pm 31$ \\ 
a0576 & 0.0389 & $4.02^{0.07}_{0.07}$ & 35 & $6.49 \pm 0.92$ & $-24.79 \pm 0.36$ & $4.62 \pm 0.06$ & $74 \pm 14$ & 	$7.55 \pm 0.04$ & $143 \pm 21$ \\ 
a3376 & 0.0456 & $4^{0.4}_{0.4}$ & 27 & $9.51 \pm 1.41$ & $-24.20 \pm 0.39$ & $3.81 \pm 0.04$ & $85 \pm 17$ & 	$5.02 \pm 0.04$ & $117 \pm 25$ \\ 
\tableline
\end{tabular}
}
\end{center}
\vskip-20pt
\end{table*}

\clearpage

\begin{table*}[htb]
\begin{center}
\tablenum{2}
\caption{Derived Quantities of the Clusters (Continued)}
{\scriptsize
\begin{tabular}{lclllcclcl}
\tableline \tableline
      &      & $T_X$ &            &  $\phi_{*,500}$              & $M_{*}-5\log h_{70}$ &  $L_{500}$                      &  & $L_{200}$ & \\
Name  & $z$  & (keV) &  $N_{obs}$ &($h_{70}^3\,$Mpc$^{-3}$)& (mag) &($h_{70}^{-2}\,10^{12} L_\odot$) & $N_{500}$ & ($h_{70}^{-2}\,10^{12} L_\odot$) & $N_{200}$ \\
  (1) & (2)  & (3)   &  (4)       & (5)                    & (6)   & (7)                             & (8)  & (9) & (10)  \\
\tableline
a0133 & 0.0569 & $3.97^{0.17}_{0.17}$ & 15 & $13.16 \pm 2.63$ & $-23.81 \pm 0.51$ & $3.72 \pm 0.04$ & $96 \pm 24$ & 	$5.35 \pm 0.03$ & $150 \pm 40$ \\ 
a0496 & 0.0328 & $3.91^{0.04}_{0.04}$ & 37 & $6.42 \pm 0.94$ & $-24.62 \pm 0.37$ & $3.91 \pm 0.05$ & $66 \pm 13$ & 	$5.47 \pm 0.04$ & $106 \pm 18$ \\ 
a1185 & 0.0325 & $3.9^{2}_{1.1}$ & 28 & $4.74 \pm 0.87$ & $-24.63 \pm 0.45$ & $2.84 \pm 0.07$ & $49 \pm 12$ & 	$4.13 \pm 0.06$ & $76 \pm 17$ \\ 
awm7 & 0.0172 & $3.9^{0.12}_{0.12}$ & 51 & $4.96 \pm 0.79$ & $-24.49 \pm 0.48$ & $2.90 \pm 0.06$ & $50 \pm 10$ & 	$3.62 \pm 0.06$ & $74 \pm 15$ \\ 
a2440 & 0.0904 & $3.88^{0.16}_{0.14}$ & 5 & $2.26 \pm 0.75$ & $-25.64 \pm 0.88$ & $3.35 \pm 0.33$ & $30 \pm 12$ & 	$4.51 \pm 0.27$ & $45 \pm 19$ \\ 
a3560 & 0.0489 & $3.87^{0.22}_{0.19}$ & 24 & $3.12 \pm 0.56$ & $-26.17 \pm 0.44$ & $6.50 \pm 0.27$ & $50 \pm 11$ & 	$9.52 \pm 0.18$ & $82 \pm 15$ \\ 
a0780 & 0.0538 & $3.8^{0.12}_{0.12}$ & 11 & $24.14 \pm 7.35$ & $-23.16 \pm 0.61$ & $3.09 \pm 0.02$ & $117 \pm 37$ & 	$3.24 \pm 0.04$ & $90 \pm 35$ \\ 
a3562 & 0.0499 & $3.8^{0.5}_{0.5}$ & 20 & $5.74 \pm 1.02$ & $-24.81 \pm 0.45$ & $3.23 \pm 0.08$ & $59 \pm 14$ & 	$6.40 \pm 0.05$ & $118 \pm 22$ \\ 
a2670 & 0.0762 & $3.73^{0.17}_{0.13}$ & 13 & $13.31 \pm 2.18$ & $-24.41 \pm 0.51$ & $5.74 \pm 0.06$ & $111 \pm 23$ & 	$8.72 \pm 0.05$ & $174 \pm 36$ \\ 
a2657 & 0.0404 & $3.7^{0.3}_{0.3}$ & 17 & $4.07 \pm 0.90$ & $-24.73 \pm 0.55$ & $2.35 \pm 0.09$ & $40 \pm 11$ & 	$2.84 \pm 0.09$ & $52 \pm 16$ \\ 
a1142 & 0.0349 & $3.7^{1.7}_{1.1}$ & 13 & $2.25 \pm 0.70$ & $-24.88 \pm 0.77$ & $1.84 \pm 0.14$ & $24 \pm 9$ & 	$2.71 \pm 0.16$ & $37 \pm 14$ \\ 
a2634 & 0.0314 & $3.7^{0.18}_{0.18}$ & 38 & $5.62 \pm 0.83$ & $-25.00 \pm 0.38$ & $4.53 \pm 0.08$ & $61 \pm 12$ & 	$6.62 \pm 0.06$ & $104 \pm 17$ \\ 
2a0335 & 0.0349 & $3.64^{0.054}_{0.048}$ & 28 & $5.81 \pm 0.95$ & $-24.72 \pm 0.41$ & $3.44 \pm 0.07$ & $56 \pm 12$ & 	$5.29 \pm 0.07$ & $86 \pm 16$ \\ 
a3526 & 0.0114 & $3.54^{0.08}_{0.08}$ & 59 & $4.02 \pm 0.71$ & $-25.12 \pm 0.67$ & $3.66 \pm 0.16$ & $44 \pm 10$ & 	$5.24 \pm 0.18$ & $68 \pm 14$ \\ 
a1367 & 0.0216 & $3.5^{0.11}_{0.11}$ & 60 & $7.81 \pm 0.99$ & $-24.62 \pm 0.34$ & $3.81 \pm 0.05$ & $69 \pm 12$ & 	$5.46 \pm 0.06$ & $93 \pm 15$ \\ 
mkw03s & 0.0434 & $3.5^{0.12}_{0.12}$ & 13 & $4.28 \pm 1.08$ & $-24.53 \pm 0.62$ & $1.96 \pm 0.08$ & $36 \pm 12$ & 	$2.76 \pm 0.06$ & $61 \pm 19$ \\ 
a1736 & 0.0458 & $3.5^{0.4}_{0.4}$ & 27 & $5.84 \pm 0.93$ & $-25.22 \pm 0.40$ & $5.28 \pm 0.10$ & $62 \pm 13$ & 	$10.97 \pm 0.08$ & $127 \pm 18$ \\ 
hcg094 & 0.0417 & $3.45^{0.3}_{0.3}$ & 12 & $3.35 \pm 0.94$ & $-24.68 \pm 0.69$ & $2.51 \pm 0.11$ & $29 \pm 10$ & 	$4.37 \pm 0.12$ & $53 \pm 14$ \\ 
a2589 & 0.0416 & $3.38^{0.08}_{0.08}$ & 18 & $6.31 \pm 1.32$ & $-24.33 \pm 0.52$ & $2.51 \pm 0.06$ & $46 \pm 13$ & 	$2.91 \pm 0.05$ & $65 \pm 20$ \\ 
mkw08 & 0.0270 & $3.29^{0.23}_{0.22}$ & 24 & $4.25 \pm 0.91$ & $-24.64 \pm 0.56$ & $2.13 \pm 0.08$ & $34 \pm 9$ & 	$2.69 \pm 0.09$ & $46 \pm 13$ \\ 
a4038 & 0.0283 & $3.15^{0.06}_{0.06}$ & 34 & $7.07 \pm 1.13$ & $-24.58 \pm 0.41$ & $2.85 \pm 0.06$ & $52 \pm 11$ & 	$3.55 \pm 0.06$ & $70 \pm 15$ \\ 
a1060 & 0.0114 & $3.1^{0.15}_{0.15}$ & 69 & $7.18 \pm 1.07$ & $-24.41 \pm 0.52$ & $2.39 \pm 0.06$ & $49 \pm 10$ & 	$3.17 \pm 0.08$ & $68 \pm 14$ \\ 
a2052 & 0.0348 & $3.1^{0.2}_{0.2}$ & 30 & $9.07 \pm 1.39$ & $-24.41 \pm 0.39$ & $3.22 \pm 0.05$ & $60 \pm 12$ & 	$4.43 \pm 0.04$ & $93 \pm 18$ \\ 
a0548e & 0.0395 & $3.1^{0.1}_{0.1}$ & 30 & $7.83 \pm 1.18$ & $-24.94 \pm 0.38$ & $4.17 \pm 0.08$ & $63 \pm 12$ & 	$6.70 \pm 0.06$ & $107 \pm 17$ \\ 
a2593 & 0.0433 & $3.1^{1.5}_{0.9}$ & 22 & $7.22 \pm 1.27$ & $-24.72 \pm 0.45$ & $3.38 \pm 0.07$ & $53 \pm 12$ & 	$5.41 \pm 0.06$ & $90 \pm 17$ \\ 
a0539 & 0.0288 & $3.04^{0.066}_{0.06}$ & 44 & $9.95 \pm 1.31$ & $-24.57 \pm 0.33$ & $3.72 \pm 0.05$ & $68 \pm 12$ & 	$4.98 \pm 0.04$ & $101 \pm 16$ \\ 
as1101 & 0.0580 & $3^{1.2}_{0.7}$ & 8 & $7.18 \pm 2.01$ & $-24.28 \pm 0.72$ & $2.18 \pm 0.08$ & $41 \pm 15$ & 	$3.16 \pm 0.05$ & $75 \pm 27$ \\ 
a0779 & 0.0230 & $2.97^{0.24}_{0.24}$ & 15 & $2.71 \pm 0.85$ & $-24.69 \pm 0.88$ & $1.72 \pm 0.14$ & $19 \pm 8$ & 	$2.25 \pm 0.19$ & $27 \pm 11$ \\ 
awm4 & 0.0326 & $2.92^{0.24}_{0.24}$ & 12 & $6.52 \pm 2.03$ & $-23.63 \pm 0.71$ & $1.48 \pm 0.04$ & $28 \pm 11$ & 	$1.73 \pm 0.04$ & $39 \pm 18$ \\ 
exo0422 & 0.0390 & $2.9^{0.5}_{0.4}$ & 11 & $4.63 \pm 1.35$ & $-24.36 \pm 0.71$ & $1.63 \pm 0.08$ & $27 \pm 10$ & 	$2.04 \pm 0.06$ & $42 \pm 16$ \\ 
a2626 & 0.0553 & $2.9^{2.5}_{1}$ & 15 & $8.85 \pm 1.61$ & $-24.67 \pm 0.50$ & $3.57 \pm 0.08$ & $57 \pm 14$ & 	$4.49 \pm 0.07$ & $77 \pm 20$ \\ 
zw1615 & 0.0302 & $2.9^{2.6}_{1.1}$ & 23 & $7.49 \pm 1.58$ & $-24.09 \pm 0.50$ & $1.68 \pm 0.05$ & $40 \pm 11$ & 	$2.18 \pm 0.05$ & $53 \pm 17$ \\ 
mkw09 & 0.0397 & $2.66^{0.57}_{0.57}$ & 12 & $7.71 \pm 2.08$ & $-23.95 \pm 0.65$ & $1.48 \pm 0.05$ & $33 \pm 12$ & 	$2.27 \pm 0.05$ & $54 \pm 17$ \\ 
a0194 & 0.0180 & $2.63^{0.15}_{0.15}$ & 24 & $3.91 \pm 0.97$ & $-24.84 \pm 0.73$ & $1.97 \pm 0.13$ & $24 \pm 8$ & 	$2.85 \pm 0.31$ & $29 \pm 10$ \\ 
a0168 & 0.0450 & $2.6^{1.1}_{0.6}$ & 21 & $10.45 \pm 1.74$ & $-24.60 \pm 0.44$ & $3.17 \pm 0.06$ & $55 \pm 13$ & 	$4.25 \pm 0.05$ & $84 \pm 18$ \\ 
a0400 & 0.0240 & $2.43^{0.078}_{0.072}$ & 29 & $7.82 \pm 1.40$ & $-24.57 \pm 0.46$ & $2.34 \pm 0.06$ & $38 \pm 9$ & 	$3.46 \pm 0.05$ & $65 \pm 13$ \\ 
a0262 & 0.0161 & $2.41^{0.03}_{0.03}$ & 39 & $7.52 \pm 1.33$ & $-24.53 \pm 0.51$ & $2.07 \pm 0.07$ & $36 \pm 8$ & 	$3.03 \pm 0.06$ & $60 \pm 12$ \\ 
a2151 & 0.0369 & $2.4^{0.06}_{0.06}$ & 21 & $6.04 \pm 1.18$ & $-25.38 \pm 0.48$ & $3.44 \pm 0.14$ & $38 \pm 9$ & 	$5.17 \pm 0.09$ & $70 \pm 13$ \\ 
mkw04s & 0.0283 & $2.13^{0.4}_{0.4}$ & 8 & $2.86 \pm 1.21$ & $-24.72 \pm 1.08$ & $1.36 \pm 0.17$ & $12 \pm 6$ & 	$1.59 \pm 0.10$ & $22 \pm 10$ \\ 
a3389 & 0.0265 & $2.1^{0.9}_{0.6}$ & 25 & $7.35 \pm 1.42$ & $-25.12 \pm 0.48$ & $2.81 \pm 0.11$ & $35 \pm 8$ & 	$4.09 \pm 0.08$ & $61 \pm 12$ \\ 
ivzw038 & 0.0170 & $2.07^{0.56}_{0.42}$ & 20 & $4.37 \pm 1.20$ & $-25.04 \pm 0.79$ & $1.76 \pm 0.17$ & $20 \pm 7$ & 	$2.34 \pm 0.12$ & $35 \pm 10$ \\ 
as0636 & 0.0116 & $2.06^{0.07}_{0.06}$ & 31 & $5.09 \pm 1.23$ & $-25.12 \pm 0.83$ & $1.89 \pm 0.19$ & $24 \pm 7$ & 	$2.40 \pm 0.24$ & $32 \pm 10$ \\ 
a3581 & 0.0230 & $1.83^{0.04}_{0.04}$ & 15 & $6.34 \pm 1.71$ & $-24.51 \pm 0.70$ & $1.22 \pm 0.09$ & $20 \pm 7$ & 	$1.52 \pm 0.16$ & $22 \pm 8$ \\ 
mkw04 & 0.0200 & $1.71^{0.09}_{0.09}$ & 13 & $4.01 \pm 1.32$ & $-25.23 \pm 0.90$ & $1.71 \pm 0.23$ & $15 \pm 6$ & 	$2.35 \pm 0.26$ & $22 \pm 8$ \\ 
ngc6338 & 0.0282 & $1.69^{0.16}_{0.16}$ & 15 & $9.07 \pm 2.22$ & $-24.39 \pm 0.61$ & $1.58 \pm 0.07$ & $23 \pm 7$ & 	$1.69 \pm 0.07$ & $29 \pm 10$ \\ 
a0076 & 0.0405 & $1.5^{1.1}_{0.6}$ & 15 & $11.29 \pm 2.42$ & $-25.14 \pm 0.53$ & $2.03 \pm 0.13$ & $30 \pm 8$ & 	$3.56 \pm 0.06$ & $58 \pm 13$ \\ 
ngc6329 & 0.0276 & $1.45^{0.08}_{0.08}$ & 13 & $9.46 \pm 2.49$ & $-24.47 \pm 0.66$ & $1.26 \pm 0.08$ & $20 \pm 7$ & 	$1.56 \pm 0.09$ & $26 \pm 9$ \\ 
as0805 & 0.0140 & $1.4^{0.3}_{0.3}$ & 28 & $13.67 \pm 2.85$ & $-24.07 \pm 0.57$ & $1.30 \pm 0.05$ & $23 \pm 6$ & 	$1.75 \pm 0.09$ & $30 \pm 9$ \\ 
ngc0507 & 0.0165 & $1.26^{0.07}_{0.07}$ & 21 & $9.39 \pm 2.23$ & $-25.18 \pm 0.63$ & $1.94 \pm 0.16$ & $21 \pm 6$ & 	$2.57 \pm 0.18$ & $28 \pm 8$ \\ 
ngc2563 & 0.0163 & $1.06^{0.04}_{0.04}$ & 13 & $10.50 \pm 3.33$ & $-24.18 \pm 0.85$ & $0.75 \pm 0.08$ & $12 \pm 5$ & 	$1.08 \pm 0.08$ & $21 \pm 7$ \\ 
wp23 & 0.0087 & $1^{0.6}_{0.4}$ & 7 & $2.71 \pm 2.71$ & $-26.01 \pm 2.83$ & $0.89 \pm 0.89$ & $6 \pm 6$ & 	$1.09 \pm 0.93$ & $10 \pm 10$ \\ 
ic4296 & 0.0133 & $0.95^{0.09}_{0.09}$ & 14 & $11.41 \pm 3.66$ & $-24.25 \pm 0.90$ & $1.25 \pm 0.09$ & $12 \pm 5$ & 	$1.63 \pm 0.13$ & $18 \pm 7$ \\ 
hcg062 & 0.0137 & $0.87^{0.02}_{0.02}$ & 8 & $7.15 \pm 3.42$ & $-24.43 \pm 1.37$ & $0.59 \pm 0.17$ & $7 \pm 4$ & 	$0.92 \pm 0.23$ & $12 \pm 6$ \\ 
\tableline
\vspace{-7 mm}
\tablecomments{Columns: (1) Name; (2) Redshift; (3) Emission-weighted
  mean temperature; (4) Estimated number of member galaxies observed; 
  (5) Characteristic
  number density; (6) Characteristic magnitude; (7) Total luminosity (with BCG)
  within $r_{500}$ brighter than $M_K=-21$; (8) Total number (with BCG) of 
  galaxies within $r_{500}$ brighter than $M_K=-21$; (9) \& (10) Same as (7) \& 
  (8), but within $r_{200}$.
  Uncertainties in all columns quoted at $1\sigma$ level. $M_*$ \& $\phi_*$ 
  calculated assuming $\alpha = -1.1$. 
  }
\end{tabular}
}
\end{center}
\vskip-23pt
\end{table*}

\bibliographystyle{apj}
\bibliography{cosmology,refs}

\end{document}